\documentclass[conference]{IEEEtran}
\IEEEoverridecommandlockouts
\usepackage{cite}
\usepackage{amsmath,amssymb,amsfonts}
\usepackage{algorithmic}
\usepackage{subfigure}
\usepackage{graphicx}
\usepackage{textcomp}

\usepackage{balance}

\usepackage{import}
\usepackage{tikz}
\usetikzlibrary{calc,fit,intersections,matrix,patterns,positioning,shapes.misc}

\graphicspath{{figs/}}

\def\BibTeX{{\rm B\kern-.05em{\sc i\kern-.025em b}\kern-.08em
    T\kern-.1667em\lower.7ex\hbox{E}\kern-.125emX}}
    
\begin{document}

\title{HoPP: 
Robust and Resilient Publish-Subscribe for an Information-Centric Internet of Things
}

\author{\IEEEauthorblockN{Cenk G{\"u}ndo\u{g}an}
\IEEEauthorblockA{
\textit{HAW Hamburg}\\
\{cenk.guendogan, peter.kietzmann\}}
\and
\IEEEauthorblockN{Peter Kietzmann}
\IEEEauthorblockA{
\textit{HAW Hamburg} \\
@haw-hamburg.de\hfill}
\and
\IEEEauthorblockN{Thomas C. Schmidt}
\IEEEauthorblockA{
\textit{HAW Hamburg} \\
t.schmidt@haw-hamburg.de}
\and
\IEEEauthorblockN{Matthias W{\"a}hlisch}
\IEEEauthorblockA{
\textit{FU Berlin}\\
m.waehlisch@fu-berlin.de}
}

\maketitle

\begin{abstract}

This paper revisits  NDN deployment in the IoT with a special focus on the interaction of sensors {\em and actuators}. Such scenarios require high responsiveness and limited control state at the constrained nodes. We argue that the NDN request-response pattern which prevents data push is vital for IoT networks. We contribute HoP-and-Pull (HoPP), a robust publish-subscribe scheme for typical IoT scenarios that targets IoT networks consisting of hundreds of resource constrained devices at intermittent connectivity. Our approach limits the FIB tables to a minimum and naturally supports mobility, temporary network partitioning, data aggregation and near real-time reactivity. We experimentally evaluate the protocol in a real-world deployment using the IoT-Lab testbed with varying numbers of constrained devices, each wirelessly interconnected via IEEE 802.15.4 LowPANs. Implementations are built on CCN-lite with RIOT and support experiments using various single- and multi-hop scenarios.

\end{abstract}


\section{Introduction}
\label{sec:intro}

The Internet of Things (IoT) is emerging, and  billions of new networked devices are forecasted. However, no common networking technology for the IoT has been agreed upon. Despite of a maturing IETF protocol suite, dozens of incompatible industry solutions are rolled out to meet device and network constraints, as well as application specific needs. 

Facing this huge world of mainly constrained devices, it seems worth rethinking its networking paradigm. A very loose coupling appears most appropriate between nodes that often run on battery with long sleep cycles and connect via lossy wireless links. Information-Centric Networking (ICN) \cite{adiko-sind-12,xvsft-sinr-14} decouples content provisioning from data producers in space which makes it a promising candidate. Additional decoupling in time and synchronization is desirable and attainable by a publish-subscribe layer.

Information-centric publish-subscribe networks have been proposed. PSIRP/PURSUIT \cite{lvt-psipp-10} is an early, prominent candidate. However, its central control architecture seems more suitable for an SDN-type deployment in LANs. Publish-subscribe schemes based on NDN like Content-based pub/sub \cite{cpw-cpsni-11} and COPSS \cite{cajfr-cecop-11} violate the loose coupling principle in their use of name-based routing or forwarding. Facing the current state of the art, we explore the problem of information-centric publish-subscribe for IoT networking open. 

In this paper, we take up the challenge and seek for an information-centric IoT networking solution that satisfies all challenges of real-world sensor-actuator networks and allows for an easy deployment. We base our work on NDN \cite{jstp-nnc-09} not only because of its widespread availability and implementations on IoT operating systems, but in particular because of its clean request-response scheme that prevents unwanted traffic at the constrained end nodes. We design and evaluate HoP-and-Pull (HoPP), a lean, adaptive publish-subscribe layer that strictly adheres to the NDN communication pattern. Our experimental findings on large IoT testbeds indicate that our system complies indeed to the challenging requirements of IoT use case with promising performance. In particular, reliability and resilience of HoPP largely outperforms previously advised push notifications.

The structure of this paper continues as follows. In Section~\ref{sec:use-cases}, we outline distinctive use cases  that motivate the following contributions. Section~\ref{sec:related_work} explores the problem space and discusses related concepts and work. In Section~\ref{sec:pub-sub}, we dive into the design details of our publish-subscribe scheme, including the key aspects of network partitioning and publisher mobility. Implementation and evaluations of our system are described in Section~\ref{sec:eval}. Finally, we conclude with an outlook in Section~\ref{sec:c+o}.  

\section{Deployment Considerations for IoT Use Cases}
\label{sec:use-cases}

In this section, we focus on  two use cases for the deployment of an information-centric IoT---the simple, well-known Lighting Control \cite{bgnt-sieoc-13}, and the more challenging application of an industrial Internet for Safety Control in harsh environments. 

\subsection{Lighting control}

Smart lighting control is essentially the task of setting the state of various lights  according to preconfigured scenarios in response to triggering events. The latter may be generated by plain switches, complex controllers, or by other machinery like an elevator that is transporting people to  a currently unilluminated floor. Configuring the proper light consists of turning various fixtures into selective settings. 

We revisit this basic use case, because it raises two interesting aspects of networking. First, lighting control foremost follows an actuation pattern, i.e., different signals request for immediate state changes at specific groups of fixtures. The information of turning a switch must somehow propagate to distributed ensembles of lights under soft real-time constraints. Burke et al. \cite{bgnt-sieoc-13} define authenticated Interests to push signalling to the actuator, thereby inverting the NDN request-response  pattern. We argue for preserving the NDN communication paradigm in Section \ref{sec:related_work}.

Second the deployment of names is closely related to the application logic and often more involved than accessing data directly. Lighting control may switch individually located fixtures (e.g., corridor light 5), or fixture groups (e.g., room 5, front), activate functions (e.g., fading), or integrate aspects into schemes (e.g.,  background illumination). Smart systems most likely combine lighting control features with further sensor readings  (e.g., user presence,  brightness detection) to apply  adaptive functions to varying device groups etc.. 

While authors in \cite{bgnt-sieoc-13} chose to combine  locations and applications within names  that are preconfigured by a control manager, we argue that preconfigured application groups at the device level are too static and violate the device context: IoT devices have an identity, capabilities, and sometimes a known location. Their role in varying application contexts, though, is extrinsic and requires a  coordinating function on the application level. This cannot be hard-coded in data names.

\subsection{Industrial safety networks}

Industrial safety and control systems are increasingly interconnected and often operate under harsh conditions. In this use case, we consider industrial environments with a threat of hazardous contaminant (e.g., explosive gas) that need continuous monitoring by stationary, as well as mobile sensors. In case of an emergency, immediate actions are required such as issuing local alarms, activating protective shut-downs (e.g., closing valves, halting pumps), initiating a remote recording for first responders and forensic purposes.

Typical industrial plants are widespread with sparse network coverage, so that mobile workers or machines face intermittent connectivity at scattered gateways. Some sensors and actuators are infrastructure bound, others are independent, battery-powered embedded devices (e.g., body equipment). The latter aspects resemble the challenges faced in previous DTN-work such as in mines \cite{gkrao-dcm-10}.

Like the previous, this use case relies on a fast sensor-actuator network including embedded IoT nodes. In addition, the harsh industrial environment raises the challenges of mobile, intermittently connected end nodes, and network partitioning. Still, enhanced reliability is required in the safety context. We will show in the following, how configurable data replication with dynamically generated content proxies can meet these challenges and how they combine in a lightweight system suitable for real-world deployment \cite{gkslp-inii-17}. 

\section{The Problem of Information Centric IoT Networking and Related Work}
\label{sec:related_work}

\subsection{Deployment in the constrained IoT}

Things in the IoT are often represented by small embedded controllers which possess orders of magnitude less resources (kBytes of memory, MHz CPU speed, mW of power) than regular Internet nodes, but still need to communicate using  protocols that interoperate in a shared infrastructure.

These things are commonly sensors or actuators that speak with a remote 'cloud' or talk with each other locally. The predominant communication for edge devices happens on wireless channels of low power lossy networks (LLNs) in the battery-powered world. Following the IEEE 802.15.4, BLE, or LWPAN standard, these nodes can exchange only small packets at very low rates and sleep frequently. Violating constraints quickly leads to successive overload, extreme packet losses, and may strongly degrade network operation or node availability. Repeated incidents have told that the mass of IoT nodes can be both highly threatened and a threat to the global Internet. 

\subsubsection{ICN in the IoT}

It became apparent \cite{olg-ccnte-10,bmhsw-icnie-14,RFC-7476} that ICN/NDN exhibit great potentials for the IoT. Not only allows the access of named content instead of distant nodes a much leaner and more robust implementation of a network layer, but in particular prevents the request-response pattern of NDN any overloading with data at the receiver. For a few years, it was the believe that NDN can be DoS resistant by design, until Interest-  and state-based attacks were discovered \cite{wsv-bipmc-12}. Subsequent work \cite{gtuz-ddndn-13,wsv-bdpts-13} elaborated the threats of Interest flooding and overloading FIB and PIT tables by user-generated names and content requests. This has proven difficult to mitigate \cite{sws-rcani-15} and is a particular threat to memory-constrained nodes.
In the subsequent Section \ref{sec:pub-sub}, we will show how a FIB with simple default routes can serve the IoT, and how PITs remain minimal by hop-wise content replication between nodes.

ICN deployment in the IoT has been studied with increasing intensity, touching protocol design aspects \cite{bmhsw-icnie-14,pf-britu-15,sqvcg-simab-16,abcmr-inmcd-16,mwt-tucin-16}, architecture work \cite{g-ainai-17,szsmb-avdir-17}, and practical use cases \cite{bgnt-sieoc-13,acim-icnis-15,srs-sndnt-15,gkslp-inii-17}. Emerging link-layer extensions for the wireless like TSCH turned out to be beneficial for the interaction of NDN communication patterns and channel management \cite{habsw-itpla-16}. Several implementations have become available. CCN-Lite \cite{ccn-lite} runs on RIOT \cite{bhgws-rotoi-13,bmhsw-icnie-14} and on Contiki \cite{dgv-clfos-04,alw-defsc-16}, NDN has been ported to RIOT \cite{saz-dinps-16}. Thus, grounds seem to be  prepared for opening the floor to real-world IoT applications with NDN.

Many deployments in the IoT, though,  follow the communication patterns {\em on demand}, {\em scheduled}, and {\em unscheduled}. Actuators in particular rely on unscheduled control messages. Since NDN is built on the request-response scheme of data-follows-Interest, unscheduled push message are not natively supported. For the IoT, this has been identified as a major research challenge \cite{RFC-7927}.

\subsubsection{Push communication} 

Several extensions have been proposed to enable an unsolicited push of data, among them  {\em Interest-follows-Interest} \cite{bgnt-sieoc-13},  {\em Interest notification} \cite{acim-ndnia-14}, and a dedicated  {\em push packet} \cite{draft-ravi-icnrg-ccn-notification}. All these push messages are sent immediately to a prospective consumer node, which not only conflicts with the ICN paradigm of naming content instead of hosts, but has no forwarding supported on the network layer. No push packet will reach its destination unless potential receivers are announced to the routing using a node-centric name. Unidirectional data push to named nodes, however, lacks flow as well as congestion control, and opens an attack surface to DoS. In the IoT with its constrained nodes, this must be rated a particularly severe disadvantage.

Carzaniga et al. \cite{cpw-cpsni-11} with a proposal of  {\em long-lived Interest} seem to be the first in addressing the push challenge in a natural NDN fashion. Subscribers issue a persistent Interest that is not consumed at content arrival, and thereby establish a (static) data path from the producer. Unfortunately, long-lived Interests open an unrestricted data path to the recipient and thereby inherit the threats of overload as other push primitives. In addition, persistent forwarding states in PITs lead to self-reinforcing broadcast storms whenever L2 broadcasts are used \cite{kgshw-nnmam-17}.
Finally, frequent topology changes as characteristic for the IoT will routinely break paths. In the following, we will show  how regular Interests with appropriate lifetime can serve this purpose equally well, without suffering from its drawbacks.

\subsubsection{The role of a control plane}

Lessons from Internet decades have told that the networking layer should be composed of well defined and clearly separated control and data planes. NDN has primarily focused on a stateful forwarding plane. 
We argue that the ICN community has payed too little attention on clearly  separating a control plane \cite{wsv-lpwds-13}.

Current proposals of routing protocols that fill NDN FIBs mainly rely on brodcasts, and often misuse Interest messages of the forwarding plane to disseminate control information. The distance vector content routing protocol DCR \cite{g-ncric-14} and the link state content routing protocol LSCR \cite{hg-nanlr-15} use broadcast pushs  to
distribute control traffic over multiple hops. This flooding is controlled by
utilizing sequence numbers and anchor nodes that store copies of the content.
An approach to reducing traffic overhead by scoped-flooding is outlined in Pro-Diluvian \cite{wboks-puscd-15}.

The set synchronization protocols ChronoSync \cite{za-lcdds-13}, iSync
\cite{fbc-snibf-15} and PartialSync \cite{zlw-pespn-16} rely on a broadcast-pull
pattern, where an Interest message containing name information is distributed
into a broadcast domain and served by the first node that maintains
conflicting name information. NLSR \cite{haazz-nnlsr-13} is a link state routing
protocol that uses ChronoSync to distribute link  information in the same
manner.

Panini \cite{swbw-lcnhp-16} explicitly defines a unicast name advertisement message (NAM) on the control plane which we will re-use when designing the publish-subscribe scheme in Section \ref{sec:pub-sub}.

\subsection{Naming and routing}

Naming content on an information-centric network layer promises a simplified access to information. Routing on names directly designs a lean network without further address mapping. It obsoletes infrastructure like the DNS and eliminates the attack surface inherent to the mapping. Both aspects are of great advantage in a constrained IoT network.
However, name-based routing encounters the problems of (a) exploding routing tables, as the number of names largely exceeds common routing resources, and (b) limited aggregation potentials, as names are specific to appliances and applications, but  independent of content locations. More severely and in contrast to IP, a local router cannot decide on aggregating names since the symbol space of names is not enumerable in practice \cite{swbw-lcnhp-16}. Limiting the complexity of name-based routing and FIB table state is one of the major challenges in IoT networks \cite{RFC-7927}.

\subsubsection{Naming in context}
\label{sec:naming+context}

In a typical IoT scenario, there are sensor readings that are reported to a  (remote) cloud, or to a  controller that operates actuators. In some cases (s. Section \ref{sec:use-cases}), sensors are co-located with a controller that generates control information for immediate actuation---a safety alert for example after a sensor threshold was exceeded.
 
Names need to be shared between the sending and the receiving side so that requests can be issued. Advertising all names throughout the routing system is infeasible and will quickly explode the FIBs. However, there are ways to mitigate this. An application-specific common knowledge, or standard naming schemes  for sensor data \cite{draft-ietf-core-senml} and alerts may obsolete the need to distribute every name to the FIBs. More generally, named {\em topics} serve as the common link in publish-subscribe systems.


In a sense, this natural approach relates to an old discussion about accessing named information in Hypermedia. Before the invention of the Worl Wide Web, Landow  \cite{l-rhm-89} already pointed out that information exchange always carries two contexts, the context of departure and that of arrival. Departure and arrival translate to publish and subscribe in our discussion.  
  
\subsubsection{Name-based routing, forwarding, and caching}

Routing normally proceeds according to location information from the FIB. Names in FIBs  only aggregate well if naming follows the topological hierarchy of the network. This rarely holds, since  naming is application-specific, and cannot be detected without distributed knowledge.  To overcome FIB explosion, several authors refer to the NDN capabilities of stateful forwarding, using the option of distributing requests  to several interfaces simultaneously \cite{yamwz-csfp-12,yaawz-rrndn-14}. Such Interest multicasting will  lead to duplicate content deliveries if the network is densely connected. In 'Pro Diluvian' \cite{wboks-puscd-15}, the effects of such scoped flooding are analyzed, and authors find a utility limited over very few ($\approx$\ 2--3) hops. Such opportunistic forwarding can also lead to loops, as was pointed out by Garcia-Luna-Aceves \cite{gm-lfpcn-16}. In any case, the excessive traffic, as well as redundant PIT states make this approach infeasible for the IoT. 

COPSS \cite{cajfr-cecop-11}, an earlier publish-subscribe approach inspired by PIM \cite{RFC-4601} multicast routing, selects a rendezvous point to interconnect publishers and subscribers. Such dedicated routing point naturally  allows for name aggregation. Like PIM-SM (Phase 2), COPSS further establishes  a dedicated forwarding infrastructure (subscription table) that establishes persistent forwarding paths from the publisher via the rendezvous point to the receivers. PANINI  \cite{swbw-panii-15,swbw-lcnhp-16} re-uses the idea of an aggregation point called Name Collector, but does not establish a (persistent) forwarding plane like COPSS. Instead, PANINI uses selective broadcasts to discover unpopular routes towards the network edge. For the IoT, we want to minimize control traffic and  avoid flooding. We restrict our solution  to a lean default routing, instead.

The ICN support of data replication and caching is of particular interest for the IoT, where wireless channels are lossy and nodes are often asleep. Hop-wise data transport with intermediate storage of chunks is a built-in feature of NDN which we extend to account for node heterogeneity. IoT deployments often consist of very constrained nodes at the edge with more powerful border routers, gateways, or other node infrastructure---many of them equipped with larger hardware, electrical connectivity, and network uplinks. In the following, we will make use of Content Proxy nodes, which are meant to be chosen from this kind. 

\subsubsection{Mobility and network partitioning}

Mobile nodes are part of many IoT deployments. While mobility is natively supported at the receiver side of NDN, publisher mobility is considered difficult to solve in a generic way \cite{tscrm-smin-13}. Translated to IoT use cases, this means mobile sensors are hard to integrate---a particular problem for surveillance and safety sensing applications. These use cases may also experience temporary network partitioning (see Section \ref{sec:use-cases}), which can be treated with correspondence to network mobility.
  
Several solutions have been built for specific applications \cite{wakvw-rtidu-12,gpwpv-hpvin-13}, but the complexity of the name-based routing system often withstands a generic mobility management. We will show in the following how prevalent default routes can naturally accommodate publisher mobility, as well as network partitioning. 


\section{HoP and Pull: A Publish-Subscribe Approach to Lightweight Routing on Names}
\label{sec:pub-sub}

\subsection{Overview}

We are now ready to describe HoP-and-Pull (HoPP), our pub-sub system for lightweight IoT deployment in detail. For a confined IoT environment, we make the common assumption that nodes form a stub network that may be connected to the outside by one or several gateways.
Some global prefix is given to a gateway, but (wireless) IoT nodes can reach a gateway without global prefix changes in one or several hops unless they are temporarily disconnected. Internally, nodes may be grouped according to one or several sub-network prefixes (e.g., {\tt /lighting}). 

We select one or several distinguished nodes to serve as Content Proxies (CPs). CPs are typically more stable and more powerful such as gateways or other infrastructural entities. These Content Proxies take the role of data caches and persistent access points. They will be reachable throughout the network by default routes, unless temporary partitioning occurs. Note that one CP can serve several local prefixes, but a local prefix may also belong to several CPs. The latter scenario will lead to replicated caching with higher and faster data availability.

Our publish-subscribe protocol for the IoT is then composed of three core primitives:

\begin{enumerate}
\item Establishing and maintaining the routing system
\item Publishing content to the CPs
\item Subscribing content from the CPs
\end{enumerate} 

Our following protocol definition strictly complies with the design principles: (a) minimal FIBs that only contain default routes, (b) no push primitive or polling, (c) no broadcast or flooding on the data plane.

\subsection{Prefix-specific default routing}
\label{sec:default_routing}

Content Proxies advertise the prefix(es) they own on the control plane to all neighbors in a Prefix Advertisement Message (PAM). Observing nodes will adopt a CP as their parent and re-broadcast the PAM message with an increased distance value. Much like in the core RPL \cite{RFC-6550}, all nodes will be members of a Destination-Oriented Directed Acyclic Graph (DODAG) after routing convergence. Nodes will include the selected best uplink in their FIB as default route to the announced prefix, but may add additional uplinks with lower priority.

\begin{figure}[h]
    \centering
    \begin{tikzpicture}[font=\sffamily,>=latex]
    \input{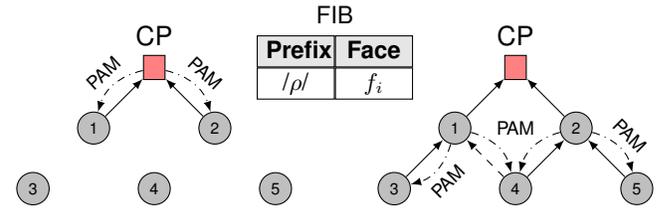}

    \pic (A) at (0,0) {tree_building=A};
    \pic[right=4.5cm of A-CP] (B) {tree_building=B};

    \path[->,dashdotted,font=\scriptsize\sffamily]
    (A-CP) edge[bend right] node[above, sloped] {PAM} (A-N1)
    (A-CP) edge[bend left] node[above, sloped] {PAM} (A-N2)
    (B-N1) edge[bend left] node[below, sloped] {PAM} (B-N3)
    (B-N1) edge[bend left] (B-N4)
    (B-N2) edge[bend right] (B-N4)
    (B-N2) edge[bend left] node[above, sloped] {PAM} (B-N5)
    ;

    \node[font=\scriptsize\sffamily] at ($(B-N1)!0.5!(B-N2)$) {PAM};

    \path[->]
    (A-N1) edge (A-CP)
    (A-N2) edge (A-CP)
    (B-N1) edge (B-CP)
    (B-N2) edge (B-CP)
    (B-N3) edge (B-N1)
    (B-N4) edge (B-N2)
    (B-N4) edge[dashed] (B-N1)
    (B-N5) edge (B-N2)
    ;


    \matrix (fibt) [fib] at ($(A-CP)!0.5!(B-CP)$)
    {
        \node[fibth]{\textbf{Prefix}}; & \node[fibth]{\textbf{Face}}; \\
        /$\rho$/ & $f_i$ \\
    };
    \node[above of=fibt, font=\small\sffamily, node distance=2em]{FIB};
\end{tikzpicture}
    \caption{Building a routing DODAG by prefix advertisements}
    \label{fig:build_dodag}
\end{figure}

Figure \ref{fig:build_dodag} visualizes the PAM prefix distribution and the corresponding FIB entry for the sample prefix /$\rho$/. All nodes establish a default on shortest paths upstream. In addition, node 4 learns a backup path of equal hop distance, but lower radio quality. 

\subsection{Publishing content}
\label{sec:pub}

An IoT node (sensor) that has new data to publish will first select a name. It may choose either  from a predefined scheme accessible by local controllers, some common standard set, or decide individually. It will advertise this content name to its upstream neighbor via a (unicast) Name Advertisement Message (NAM). It will also associate the content with one or several topic names and adds these to the content metadata.

\begin{figure*}[t]
    \centering
    \begin{tikzpicture}[font=\sffamily,>=latex]
    \input{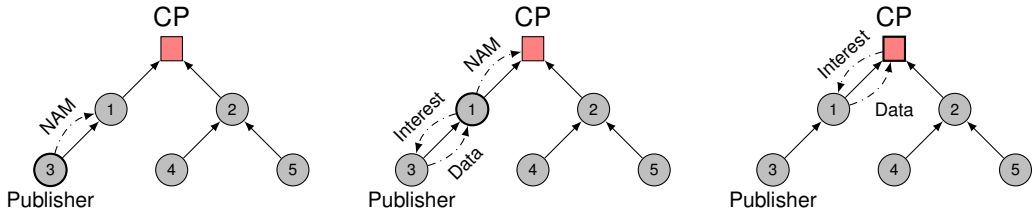}

    \pic (A) at (0,0) {tree_building=A};
    \pic[right=4.5cm of A-CP] (B) {tree_building=B};
    \pic[right=4.5cm of B-CP] (C) {tree_building=C};

    \node[font=\footnotesize\sffamily] at ([yshift=-5pt]A-N3.south) {Publisher};
    \node[font=\footnotesize\sffamily] at ([yshift=-5pt]B-N3.south) {Publisher};
    \node[font=\footnotesize\sffamily] at ([yshift=-5pt]C-N3.south) {Publisher};

    \node[font=\footnotesize\sffamily] at ([yshift=-5pt]A-N5.south) {\phantom{Subscriber}};
    \node[font=\footnotesize\sffamily] at ([yshift=-5pt]B-N5.south) {\phantom{Subscriber}};
    \node[font=\footnotesize\sffamily] at ([yshift=-5pt]C-N5.south) {\phantom{Subscriber}};

    \node[router,fill=none,thick] at (A-N3) {\phantom{3}};
    \node[router,fill=none,thick] at (B-N1) {\phantom{1}};
    \node[contentproxy,fill=none,thick] at (C-CP) {\phantom{0}};

    \path[->]
    (A-N1) edge (A-CP)
    (A-N2) edge (A-CP)
    (A-N3) edge (A-N1)
    (A-N4) edge (A-N2)
    (A-N5) edge (A-N2)
    (B-N1) edge (B-CP)
    (B-N2) edge (B-CP)
    (B-N3) edge (B-N1)
    (B-N4) edge (B-N2)
    (B-N5) edge (B-N2)
    (C-N1) edge (C-CP)
    (C-N2) edge (C-CP)
    (C-N3) edge (C-N1)
    (C-N4) edge (C-N2)
    (C-N5) edge (C-N2)
    ;

    \path[->,dashdotted, font=\scriptsize\sffamily, sloped]
    (A-N3) edge[bend left] node[above] {NAM} (A-N1)
    (B-N1) edge[bend left] node[above] {NAM} (B-CP)
    (B-N1) edge[bend right] node[above] {Interest} (B-N3)
    (B-N3) edge[bend right] node[below] {Data} (B-N1)
    (C-CP) edge[bend right] node[above] {Interest} (C-N1)
    (C-N1) edge[bend right] (C-CP)
    ;
    \node[font=\scriptsize\sffamily] at ($(C-N1)!0.5!(C-N2)$) {Data};

\end{tikzpicture}
    \caption{Publishing new content: Advertising names and pulling content  hop-by-hop}
    \label{fig:publish}
\end{figure*}

Under regular network conditions, the upstream neighbor is expected to retrieve the advertised content via the incoming interface of the NAM. It proceeds according to the standard NDN scheme: An Interest requests the name, the data is returned in response. Concurrently, the upstream issues a corresponding NAM to its parent, which in turn pulls the content one hop closer to the CP. This hop-wise content replication proceeds until the data arrives at the Content Proxy.  

It is worth noting that the NAM content alerting is situated on the control plane using {\em link-local unicast} signaling. Neither a data path is established in the PIT, nor are FIBs modified. Hop-wise content retrieval is also more robust to changing network conditions, while experiencing little temporal overhead when executed in parallel.    
   
The publishing mechanism is depicted in Figure~\ref{fig:publish}. Publisher~3 issues a NAM to its parent 1, which requests the content and republishes the NAM to the CP in parallel. After arrival of the data, node 1 can satisfy the Interest which was received by the CP.

Under irregular network conditions, a node may not receive an Interest that matches its previous name advertisements. This may be due to broken links, failing or deep-sleeping nodes, or enduring overload. After a deployment-specific timeout, the content owner will adapt and try to publish the content on an alternate path by sending a NAM up on a backup link. In case of a complete failure, the content node can follow two strategies: Either it waits and re-advertises  according to an exponential back-off, or it solicits a refresh of router advertisements for learning new, operational routes.

\subsection{Subscribing to content}

A subscriber in HoPP  behaves almost like any content requester in NDN. It issues a regular Interest request up the default route to the CP and awaits the response. There are two deviations from plain NDN, though. First, the subscriber cannot extract content names from its FIB, since FIBs only contain prefixes. Second, it does not expect an immediate reply, but issues Interests with extended lifetimes. 

\begin{figure*}[t]
    \centering
    \begin{tikzpicture}[font=\sffamily,>=latex]
    \input{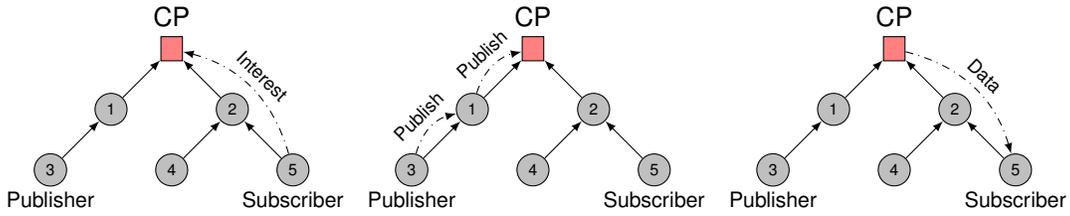}

    \pic (A) at (0,0) {tree_building=A};
    \pic[right=4.5cm of A-CP] (B) {tree_building=B};
    \pic[right=4.5cm of B-CP] (C) {tree_building=C};

    \node[font=\footnotesize\sffamily] at ([yshift=-5pt]A-N3.south) {Publisher};
    \node[font=\footnotesize\sffamily] at ([yshift=-5pt]A-N5.south) {Subscriber};
    \node[font=\footnotesize\sffamily] at ([yshift=-5pt]B-N3.south) {Publisher};
    \node[font=\footnotesize\sffamily] at ([yshift=-5pt]B-N5.south) {Subscriber};
    \node[font=\footnotesize\sffamily] at ([yshift=-5pt]C-N3.south) {Publisher};
    \node[font=\footnotesize\sffamily] at ([yshift=-5pt]C-N5.south) {Subscriber};

    \path[->]
    (A-N1) edge (A-CP)
    (A-N2) edge (A-CP)
    (A-N3) edge (A-N1)
    (A-N4) edge (A-N2)
    (A-N5) edge (A-N2)
    (B-N1) edge (B-CP)
    (B-N2) edge (B-CP)
    (B-N3) edge (B-N1)
    (B-N4) edge (B-N2)
    (B-N5) edge (B-N2)
    (C-N1) edge (C-CP)
    (C-N2) edge (C-CP)
    (C-N3) edge (C-N1)
    (C-N4) edge (C-N2)
    (C-N5) edge (C-N2)
    ;

    \path[->,dashdotted, font=\scriptsize\sffamily, sloped]
    (A-N5) edge[bend right] node[above] {Interest} (A-CP)
    (B-N3) edge[bend left] node[above] {Publish} (B-N1)
    (B-N1) edge[bend left] node[above] {Publish} (B-CP)
    (C-CP) edge[bend left] node[above] {Data} (C-N5)
    ;

\end{tikzpicture}
    \caption{Content subscription: Requesting name by topic with asynchronous delivery}
    \label{fig:subscribe}
\end{figure*}

Names are expected to follow an application-specific logic. Following up the discussion in Sections \ref{sec:naming+context}, we argue that names (of content or topics) in machine-to-machine communication must be processable in the context of the endpoint and thus known. Names of individual content items can be learned by issuing Interests on topics. The corresponding CP will then answer the request with an empty data chunk that carries available content name(s) as metadata.  

Figure \ref{fig:subscribe} displays the operations of a subscriber. An Interest for named content is sent up to the proper prefix owner (CP) and remains for a predefined lifetime, if the Content Proxy cannot supply the data. In case  content is arriving from a publisher to the CP, data is transferred automatically down the reverse Interest path---as a regular NDN operation. We anticipate that in common sensor-actuator networks of the IoT, the application semantic will define meaningful Interest lifetimes. Otherwise, in regimes of largely fluctuating temporal behaviours or long-lasting subscriptions (e.g., alerts), the subscriber may refresh and maintain the request at its discretion.

Note that in contrast to {\em long-lived Interests} or the COPSS {\em subscription tables} (s. Sec. \ref{sec:related_work}), such Interests of extended lifetime are consumed by arriving content and do not open a persistent,  uncontrolled data path. Subscribers continue to apply flow control and may discontinue subscriptions to unwanted content. 

\subsection{Publisher mobility and network partitioning}

A publishing node that moves from one point of attachment to another within the IoT domain, will experience stable routing conditions in the sense that default routes to active prefixes should exist everywhere in a connected network. Correspondingly, the mobile node (MN) can re-configure its upstream route either by wait for the next prefix advertisement (PAM), or may actively solicit an additional PAM. Note that these link-local route configurations closely resemble the autoconfiguration of IPv6 default gateways. However, in contrast to mobile IPv6, the MN in our publish-subscribe system can continue publication immediately after a link-local route is established.    

\begin{figure}[h]
    \centering
    \begin{tikzpicture}[font=\sffamily,>=latex]
    \input{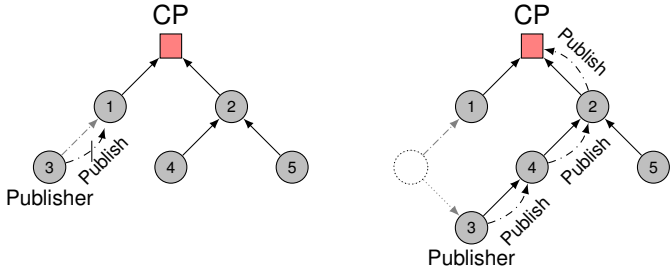}

    \pic (A) at (0,0) {tree_building=A};
    \pic[right=4.5cm of A-CP] (B) {tree_building=B};
    \node[router,scale=1.25,fill=white,draw=none] at (B-N3) {};
    \node[router,fill=white,densely dotted,scale=1.05] (old) at (B-N3) {\phantom{3}};
    \node[router, below left of=B-N4] (new) {3};
    \node[font=\footnotesize\sffamily] at ([yshift=-5pt]A-N3.south) {Publisher};
    \node[font=\footnotesize\sffamily] at ([yshift=-5pt]new.south) {Publisher};

    \node[font=\footnotesize\sffamily] at ([yshift=-5pt]A-N5.south) {\phantom{Publisher}};
    \node[font=\footnotesize\sffamily] at ([yshift=-5pt]B-N5.south) {\phantom{Publisher}};

    \path[->,dashdotted,font=\scriptsize\sffamily]
    (A-N3) edge[bend right,shorten >= 2pt]
        node[sloped,strike out,draw=black,-,solid]{}
        node[below, sloped] {Publish} (A-N1)
    (B-N4) edge[bend right] node[below, sloped] {Publish} (B-N2)
    (B-N2) edge[bend right] node[above, sloped] {Publish} (B-CP)
    (new) edge[bend right] node[below, sloped] {Publish} (B-N4)
    (old) edge[draw=gray,densely dotted] (new)
    ;

    \path[->]
    (A-N1) edge (A-CP)
    (A-N2) edge (A-CP)
    (A-N3) edge[densely dashdotted,draw=gray] (A-N1)
    (A-N4) edge (A-N2)
    (A-N5) edge (A-N2)
    (B-N1) edge (B-CP)
    (B-N2) edge (B-CP)
    (B-N4) edge (B-N2)
    (B-N3) edge[densely dashdotted,draw=gray] (B-N1)
    (B-N5) edge (B-N2)
    (new) edge (B-N4)
    ;
\end{tikzpicture}
    \caption{Publisher Mobility: Switching DODAGs}
    \label{fig:publisher_mobility}
\end{figure}

Figure \ref{fig:publisher_mobility} illustrates provider mobility. Node 3 removes from the network while trying to publish a content item and enters the radio range of node 4. It may now actively learn about network re-attachment (e.g., from link triggers), or learn from a newly arriving PAM. After the local upstream is configured, the mobile publisher can successfully complete its publishing handshake.  

\begin{figure}[h]
    \centering
    \begin{tikzpicture}[font=\sffamily,>=latex]
    \input{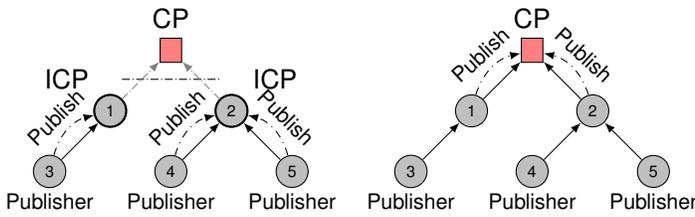}

    \pic (A) at (0,0) {tree_building=A};
    \pic[right=4.5cm of A-CP] (B) {tree_building=B};
    \node[router,fill=none,thick,label={above right:ICP}] at (A-N2) {\phantom{1}};
    \node[router,fill=none,thick,label={above left:ICP}] at (A-N1) {\phantom{2}};
    \node[font=\footnotesize\sffamily] at ([yshift=-5pt]A-N3.south) {Publisher};
    \node[font=\footnotesize\sffamily] at ([yshift=-5pt]A-N4.south) {Publisher};
    \node[font=\footnotesize\sffamily] at ([yshift=-5pt]A-N5.south) {Publisher};
    \node[font=\footnotesize\sffamily] at ([yshift=-5pt]B-N3.south) {Publisher};
    \node[font=\footnotesize\sffamily] at ([yshift=-5pt]B-N4.south) {Publisher};
    \node[font=\footnotesize\sffamily] at ([yshift=-5pt]B-N5.south) {Publisher};

    \path[->,dashdotted, font=\footnotesize\sffamily]
    (A-N4) edge[bend left] node[above, sloped] {Publish} (A-N2)
    (A-N3) edge[bend left] node[above, sloped] {Publish} (A-N1)
    (A-N5) edge[bend right] node[above, sloped] {Publish} (A-N2)
    (B-N2) edge[bend right] node[above, sloped] {Publish} (B-CP)
    (B-N1) edge[bend left] node[above, sloped] {Publish} (B-CP)
    ;

    \draw[densely dashdotted] ([yshift=7.5pt]A-N1.north east) -- ([yshift=7.5pt]A-N2.north west);

    \path[->]
    (A-N3) edge (A-N1)
    (A-N1) edge[densely dashdotted,draw=black!50] (A-CP)
    (A-N2) edge[densely dashdotted,draw=black!50] (A-CP)
    (A-N4) edge (A-N2)
    (A-N5) edge (A-N2)
    (B-N1) edge (B-CP)
    (B-N2) edge (B-CP)
    (B-N3) edge (B-N1)
    (B-N4) edge (B-N2)
    (B-N5) edge (B-N2)
    ;
\end{tikzpicture}
    \caption{Temporary network partitioning: Interim Content Proxies (ICPs) buffer publishing}
    \label{fig:partitioning}
\end{figure}

Temporary network partitioning proceeds very similar to mobility. An intermediate node that looses upstream connectivity will explore alternate paths (cf. Sec.~\ref{sec:pub}), but has to await a re-attachment in case of a complete failure. Such node will continue to  receive publishing demands (NAMs) from the downstream, which it will satisfy in accordance with its resources. On overload, it will terminate to retrieve content from its children. Proceeding this way will establish a classic backpressure mechanism of flow control.

Operations under network partitioning are shown in Figure~\ref{fig:partitioning}. Following an outage of the CP, nodes 1 and 2 experience a disconnect. They continue to handle publications (as well as subscriptions) until connectivity to the CP is reestablished.

\section{Implementation and Evaluation}
\label{sec:eval}

\subsection{Implementation for CCN-lite on RIOT}

\begin{figure}[h]
    \centering
    \resizebox{0.90\columnwidth}{!}{\import{figs/}{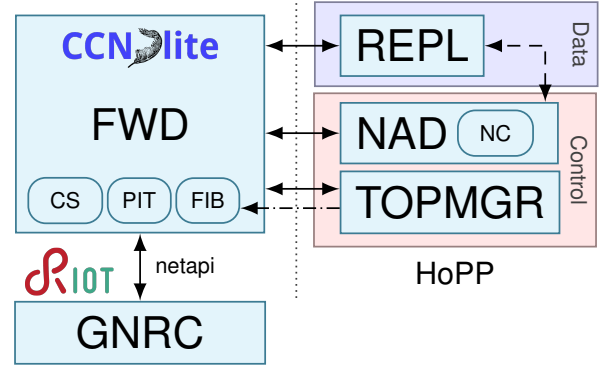}}
    \caption{IoT Publish-Subscribe Architecture}
    \label{fig:arch}
\end{figure}

We implemented the HoPP extensions on the CCN-lite version ported to RIOT and deploy NDN. It is noteworthy that this software stack supports both,  the NDN core protocol as well as CCNx.  On RIOT, CCN-lite implements the {\tt netdev} interface and  runs as a dedicated single-threaded network stack. 

The architecture of the extended CCN-lite is depicted in Figure
\ref{fig:arch}. It mainly adds a new control protocol block that 
handles exchange and processing of the two new packet types (\texttt{PAM, NAM}) on the control plane. This extends the
\texttt{forwarder} module of CCN-lite. The \texttt{forwarder} allows extensions for the
packet parsing by the use of user-defined callback functions on a suite
basis. Considering this loose coupling, the actual topology maintenance was
implemented separately from the CCN-lite core. The \texttt{topology manager}
handles \texttt{PAM} scheduling and parent selection to form and maintain the
routing topology (DODAG). Resulting forwarding states are reflected in the
FIB with the help of the CCN-lite  API. The Name Advertisement
Daemon (\texttt{NAD}) module handles parsing and scheduling of \texttt{NAM}
messages. A \texttt{NAM} Cache (\texttt{NC}) is used to intermittently track
the hop-wise propagation and to reschedule \texttt{NAM} transmissions in case
of network disruptions. For each entry in the \texttt{NC}, the \texttt{NAD}
triggers the \texttt{replicator} to invoke a hop-wise content replication
on the data plane via pull-driven Interest-Data.
To ensure hop-wise replication of published content, a caching strategy was added to CCN-lite that
hinders replicated content to be cached out during  publishing.
After a successful Interest-Data exchange, the \texttt{replicator} notifies the \texttt{NAD} module and the appropriate \texttt{NC} entry is freed for removal.


\begin{figure}
   \includegraphics[width=0.94\columnwidth]{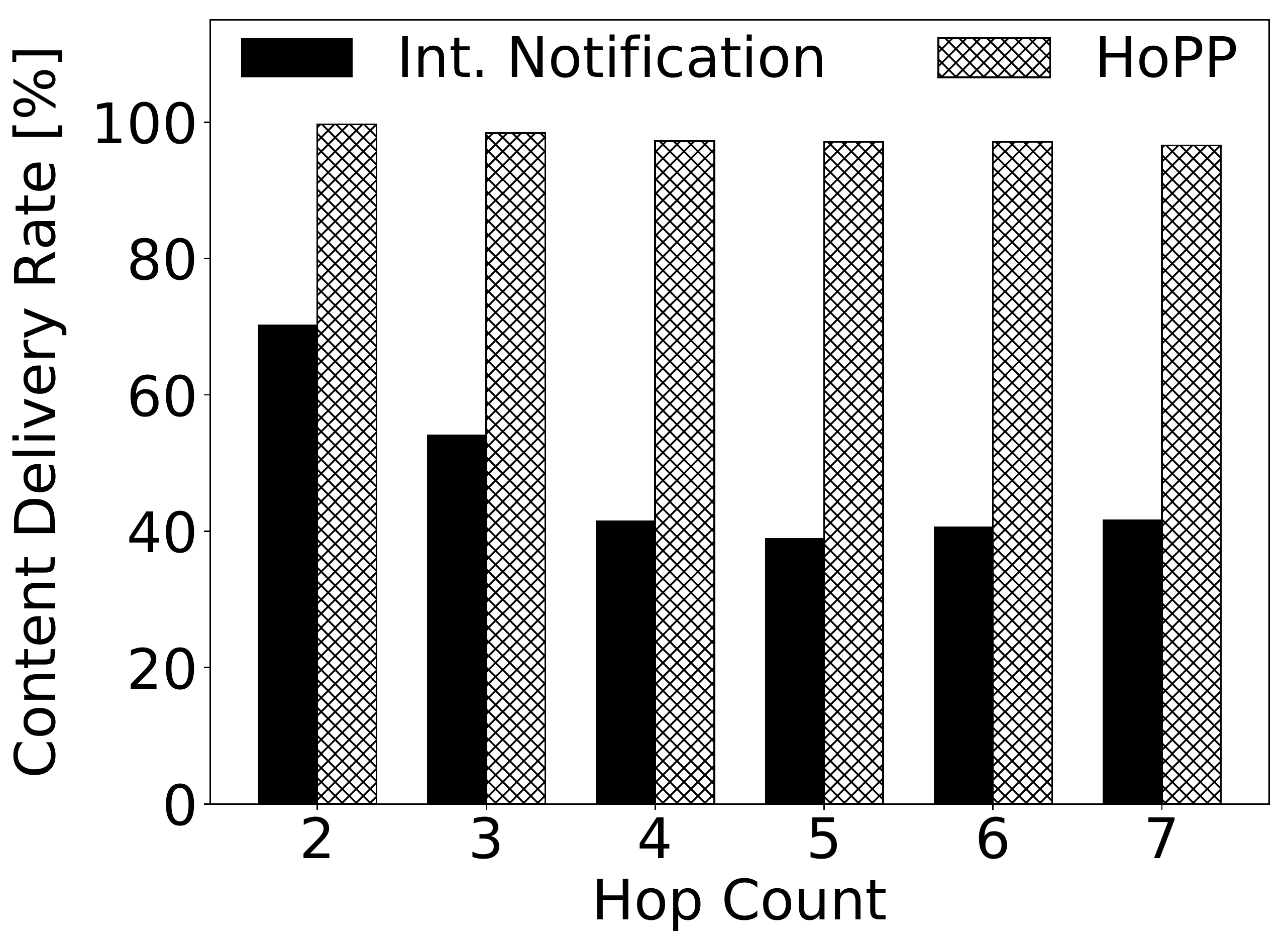}
    \caption{Success rate of content delivery to one consumer as a function of hop count}
    \label{fig:reliability}
\end{figure}

\subsection{Basic Testbed Setup}

All experiments are conducted in the FIT~IoT-LAB testbed \cite{ivlso-14} to reflect common IoT properties.
The testbed consists of several hundreds of class~2 devices equipped with an ARM Cortex-M3 MCU, 
64~kB of RAM and 512~kB of ROM, and an IEEE~802.15.4 radio (Atmel AT86RF231). 
The radio card provides basic MAC layer functions implemented in hardware, such as ACK handling, 
retransmissions, and CSMA/CA (Carrier sense multiple access/Collision avoidance).
The software platform is based on RIOT \cite{bhgws-rotoi-13} and the CCN-lite network stack \cite{ccn-lite}, including the protocol extensions described above.

The performance of the HoPP publish-subscribe IoT system is evaluated on the three different topologies: 

\begin{LaTeXdescription}
\item[Paris] is a densely connected topology of 69 nodes all within radio reach. 
\item[Grenoble (ring)] is formed of a closed rectangle with two double-stacked edges. 178 nodes form a heterogeneously meshed network with a maximal hop distance of four. 
\item[Grenoble] consists of about 350 nodes, where half of them is situated on the rectangle, the other half forms linear extensions leading outwards. This network organizes in complex, fluctuating topologies with a node distance up to 9 hops. 
\end{LaTeXdescription}

\subsection{Performance evaluation}

The first evaluation inspects the reliability of HoPP as compared to plain Interest notification. 
We investigate the content reception rate on a given consumer in the Grenoble ring multi-hop topology using a converge cast traffic pattern, where each device generates sensor readings every $30 \pm 15$~seconds.
While HoPP is able to build and maintain the topology, static forwarding states were installed on the devices for the {Interest Notification} approach using the same routing information as HoPP.

Figure \ref{fig:reliability} compares the reliability of HoPP with the common {Interest Notification} approach in relation to the hop distance of the consumer.
For HoPP, we observe a steady high content delivery rate above $96~\%$ for all hop distances in the topology.
NDN {Interest Notification}  admits significantly lower reliability and shows a decline in transmission with increasing hop distance. While a hop count of $1$ yields $~70~\%$ packet arrivals, success ratio decreases to $~41~\%$ for hop distances of $5$ and larger.
Next, we investigate performance metrics that relate to the temporal behaviour of the protocol. Since deficits of the core protocol, but also different failures of networked elements (radio/link layer, CCN-layer, pub-sub, and node layer) translate into delays due to retransmissions and re-arrangements, times to completion are a key performance indicators. In detail, we study (i) routing convergence, (ii) times to publish content items, (iii) times to publish under network partitioning,  and (iv) times to issue alerts (from publisher to the subscribers). 

\begin{figure}
   \includegraphics[width=0.94\columnwidth]{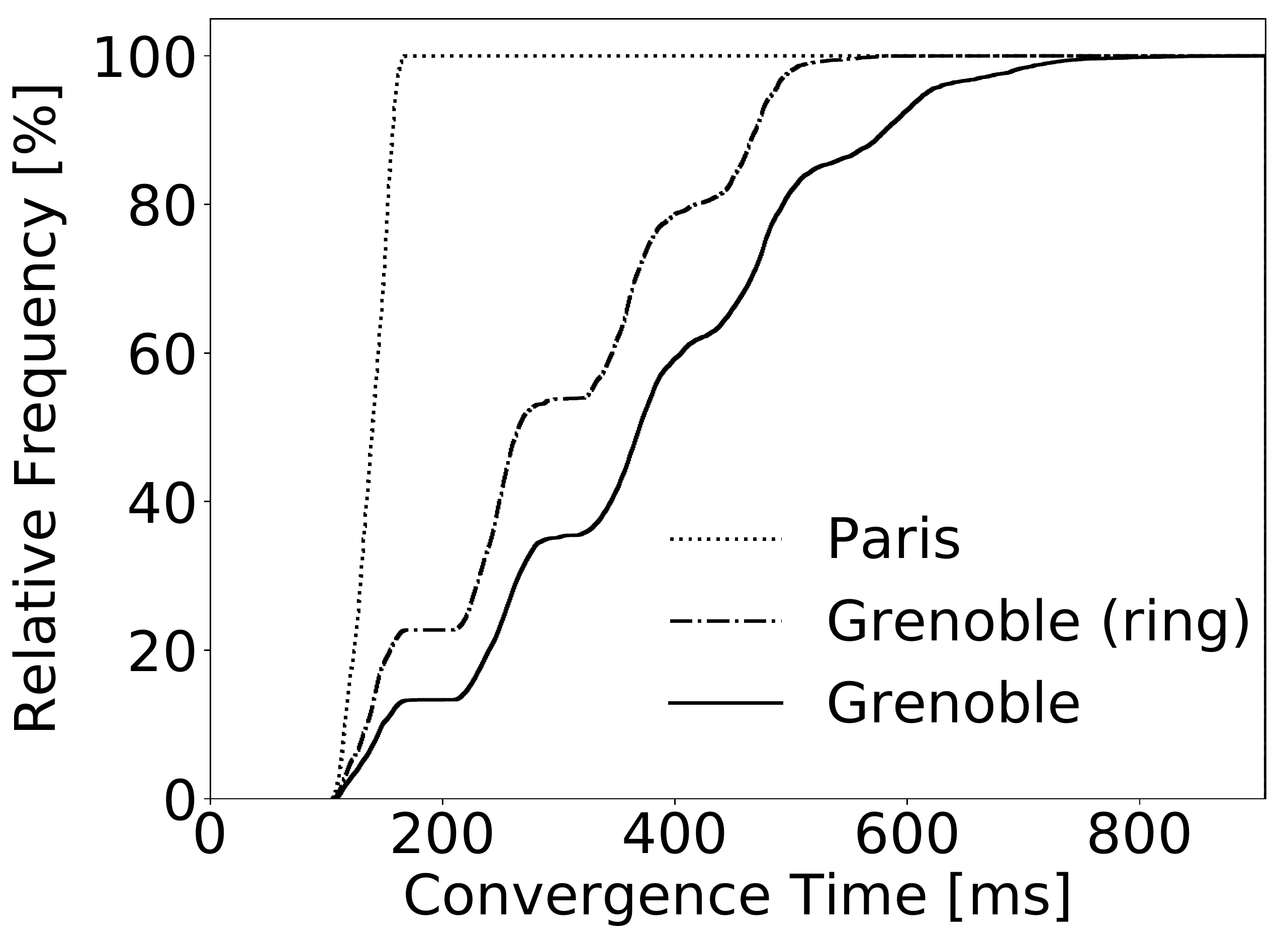}
    \caption{Routing convergence time for the testbed topologies}
    \label{fig:routing_convergence}
\end{figure} 

\begin{figure*}
   \subfigure[Paris topology]{\includegraphics[width=0.66\columnwidth]{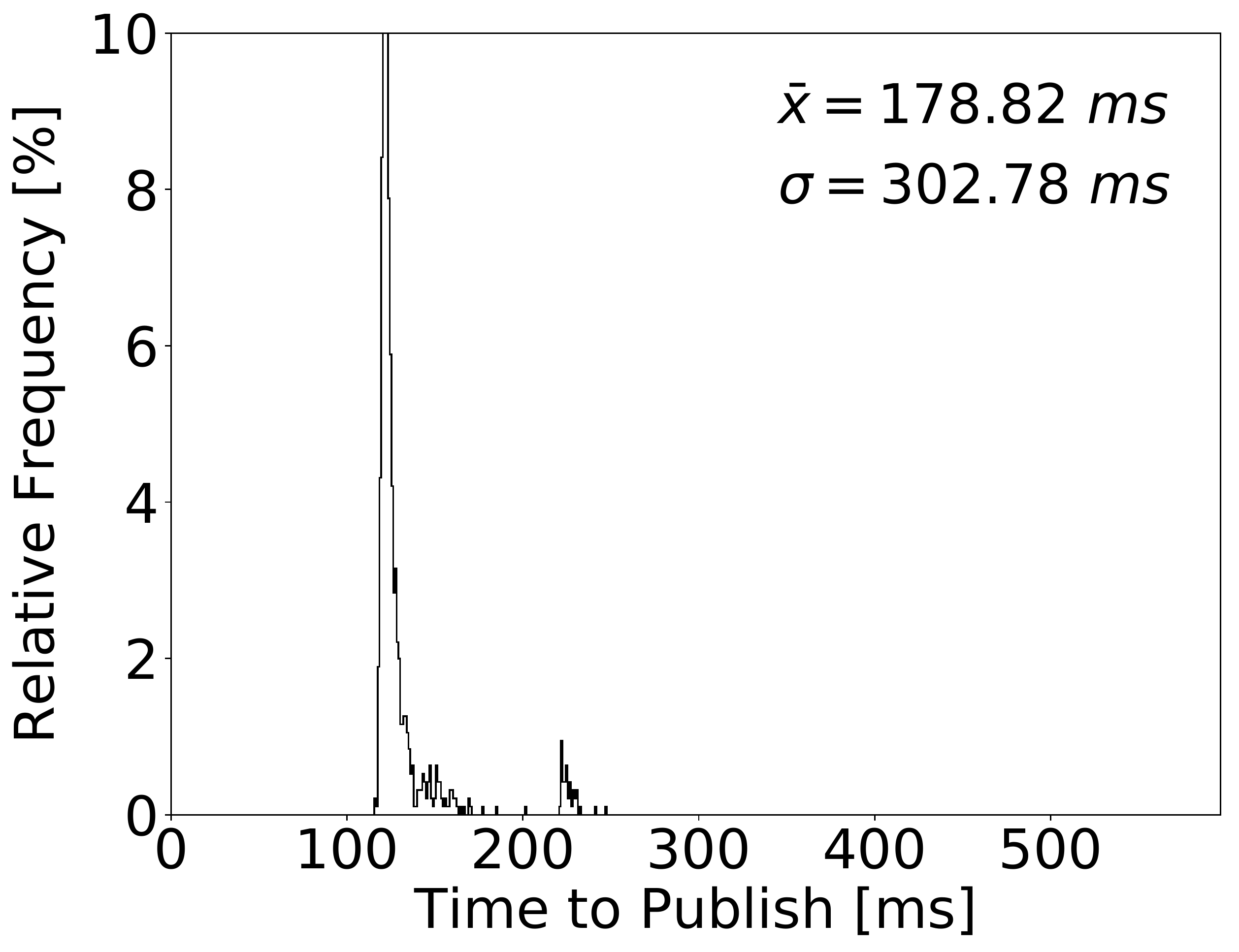}
    \label{fig:pubtime_paris}}
   \subfigure[Grenoble (ring) topology]{\includegraphics[width=0.66\columnwidth]{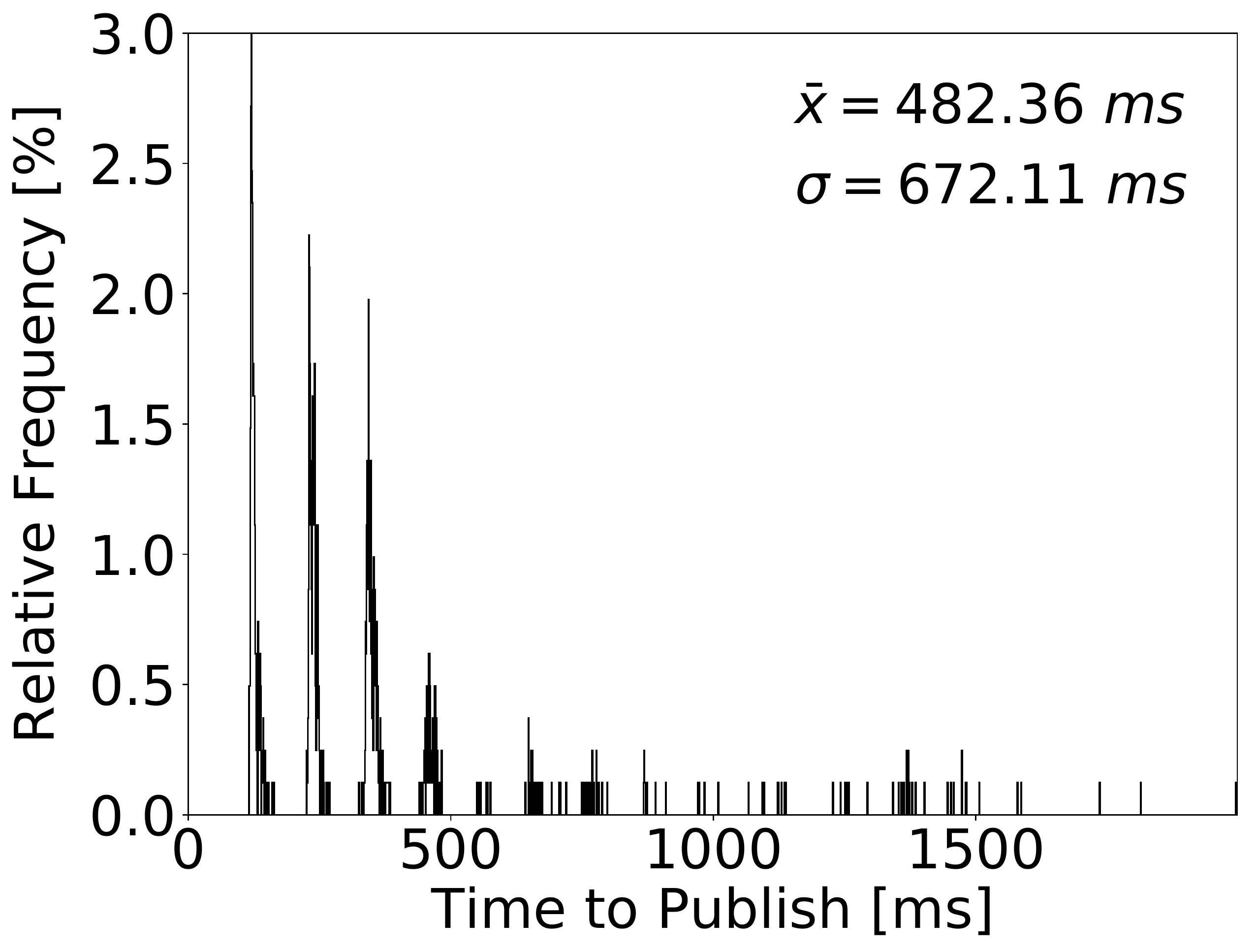}
    \label{fig:pubtime_grenoble-ring}}
   \subfigure[Grenoble topology]{\includegraphics[width=0.66\columnwidth]{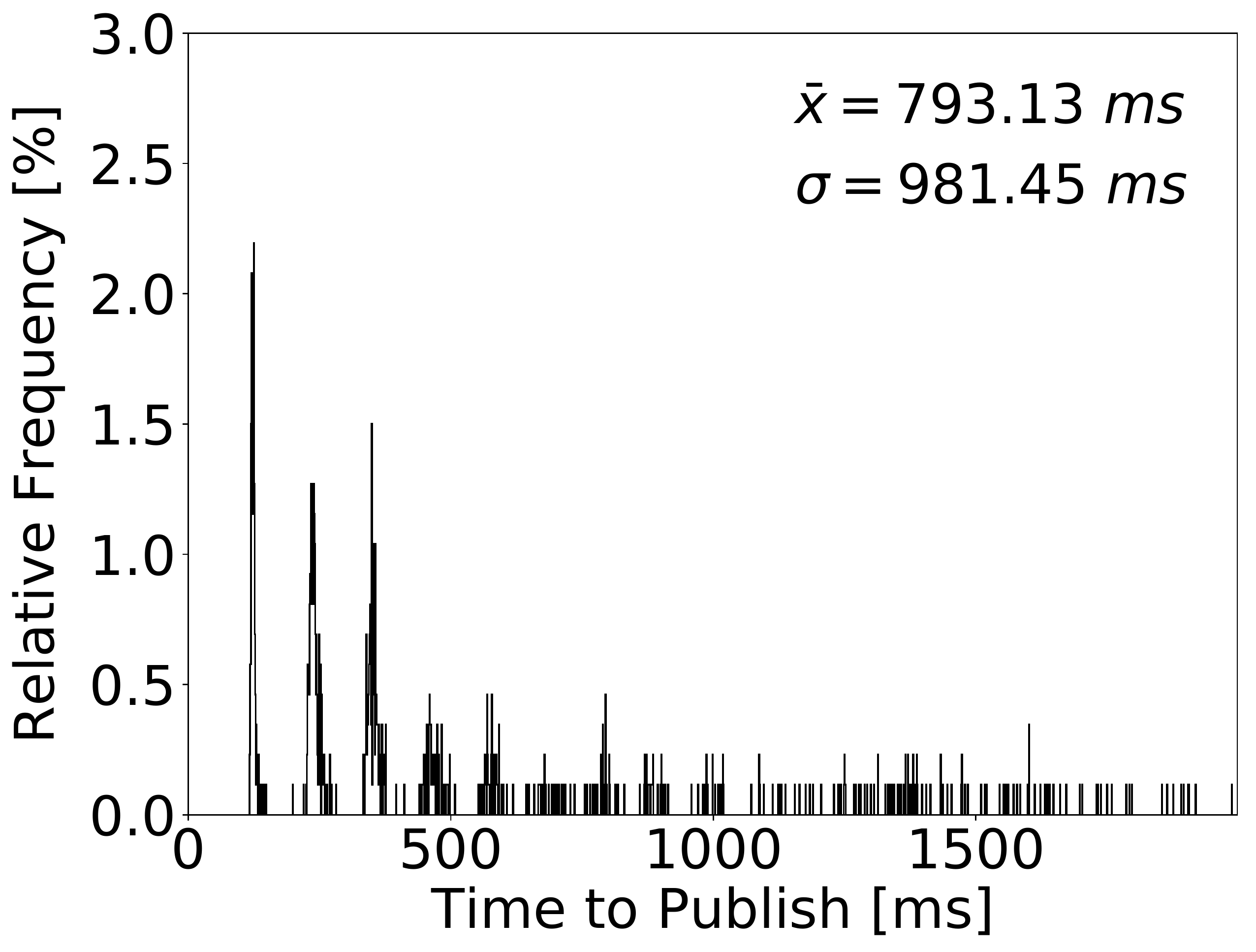}
    \label{fig:pubtime_grenoble}}
    \caption{Time to content publishing}
\end{figure*}

Routing convergence times in the three testbeds are displayed in Figure \ref{fig:routing_convergence}. Clearly visible is the dependence on hop counts, each counting for an average delay of $\approx 100$ ms---the PAM timer. While Paris is single-hop network and exhibits a single step in distribution, multiple steps represent hop count multiplicities in the multi-hop cases. No exceptional delays become visible. This is due to the moderate timing of the routing protocol which causes a low network utilization.
 
For the evaluation of the times needed to publish a content item, we iterate the following scenario. For each topology, a Content Proxy is positioned in the center of the network, while randomly chosen nodes publish a single, individually named chunk to the network. Publication is initiated every second and depending on the nodes position in the tree, one to several data packets might traverse the same sub-paths within this time frame.

Results for the single-hop network (Paris) are displayed in Figure  \ref{fig:pubtime_paris}. Observing round-trip ping values of $\approx 10$~ms, the NAM timer ($\mathtt{nam_t}$) of $125 \pm 25$~ms, and the CCN-lite processing, a mean time to publish of about $135$ ms would be expected. Small fluctuations at $\approx 2 \times \mathtt{nam_t}$ indicate additional delays that result from network disturbances and node congestion leading to paths of hop count two.

Similar results become visible from the Grenoble experiments in Figs. \ref{fig:pubtime_grenoble-ring} and \ref{fig:pubtime_grenoble}. Clearly pronounced are the first four routing hops, higher hop counts in  Fig. \ref{fig:pubtime_grenoble} blur according to increasing fluctuations. These results clearly show the fragility of the lossy wireless regime, but also confirm a majority of these challenging transmissions did complete on the expected time scale.

We analyzed a scenario of network partitioning on the Grenoble ring topology. To quantify the effects of a major network disruption, we disabled all nodes of rank two every 60 s for an off-time interval of 60 s. This isolated the Content Proxy periodically. Content publishing proceeded randomly with a frequency of one per second.  

\begin{figure}
   \includegraphics[width=0.94\columnwidth]{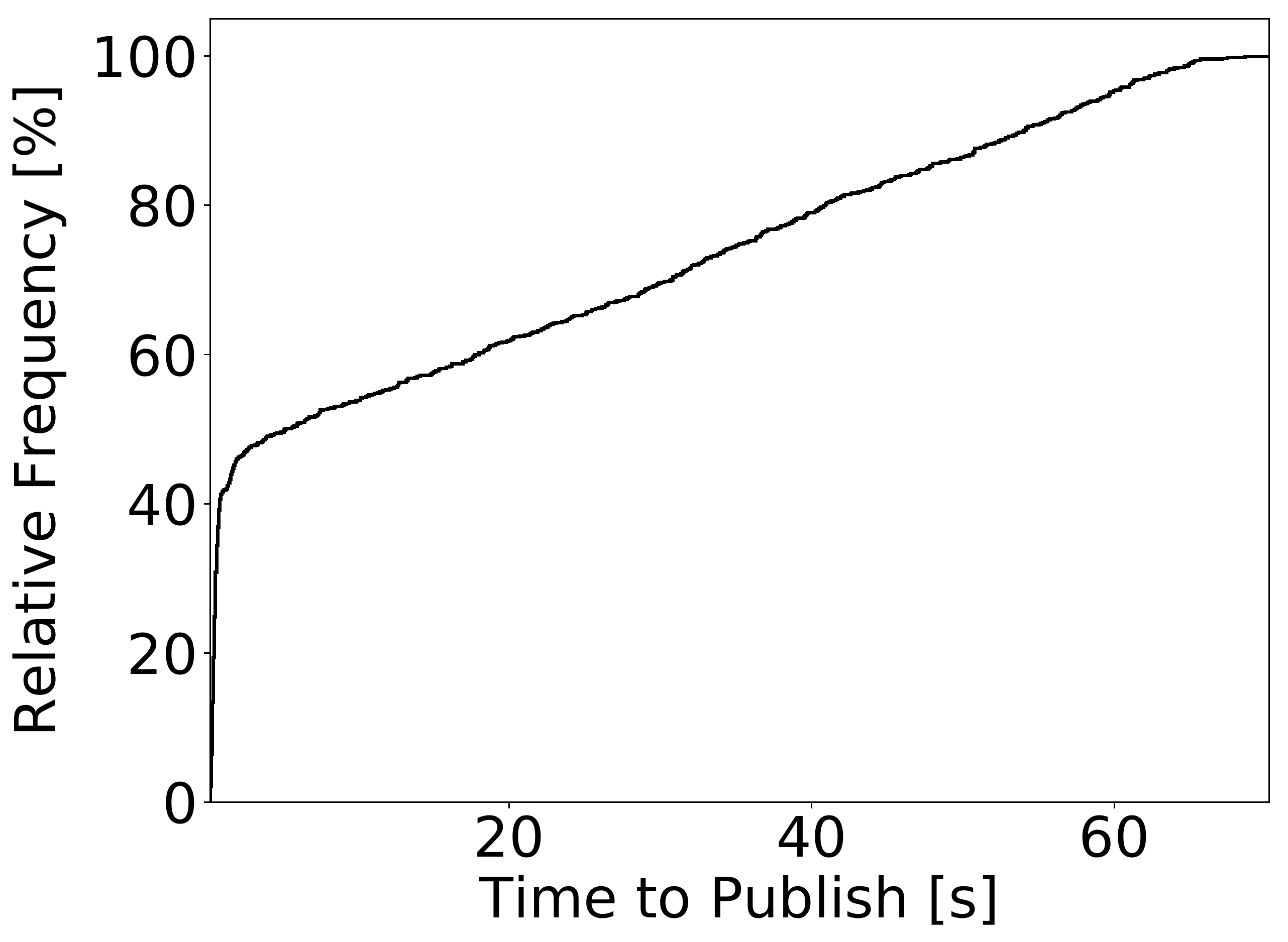}
    \caption{Time to content publishing at network partitioning}
    \label{fig:partitioning_grenoble-ring}
\end{figure} 
 
Results in Figure \ref{fig:partitioning_grenoble-ring} highlight a smooth content transition to the CP with a timing  almost linearly stretched over the 60 s off-period. No unexpected content delays become visible, which indicates the protocol robustness on this macroscopic time scale.

Finally, the end-to-end delay from the publisher to the subscriber was examined. This corresponds to the use case of issuing alerts between nodes from the local IoT network. The scenarios correspond to the previous measurements of the publishing time, i.e., publishing and subscription requests are issued randomly scattered within the topology at intervals of one second.

\begin{figure*}
   \subfigure[Paris topology]{\includegraphics[width=0.66\columnwidth]{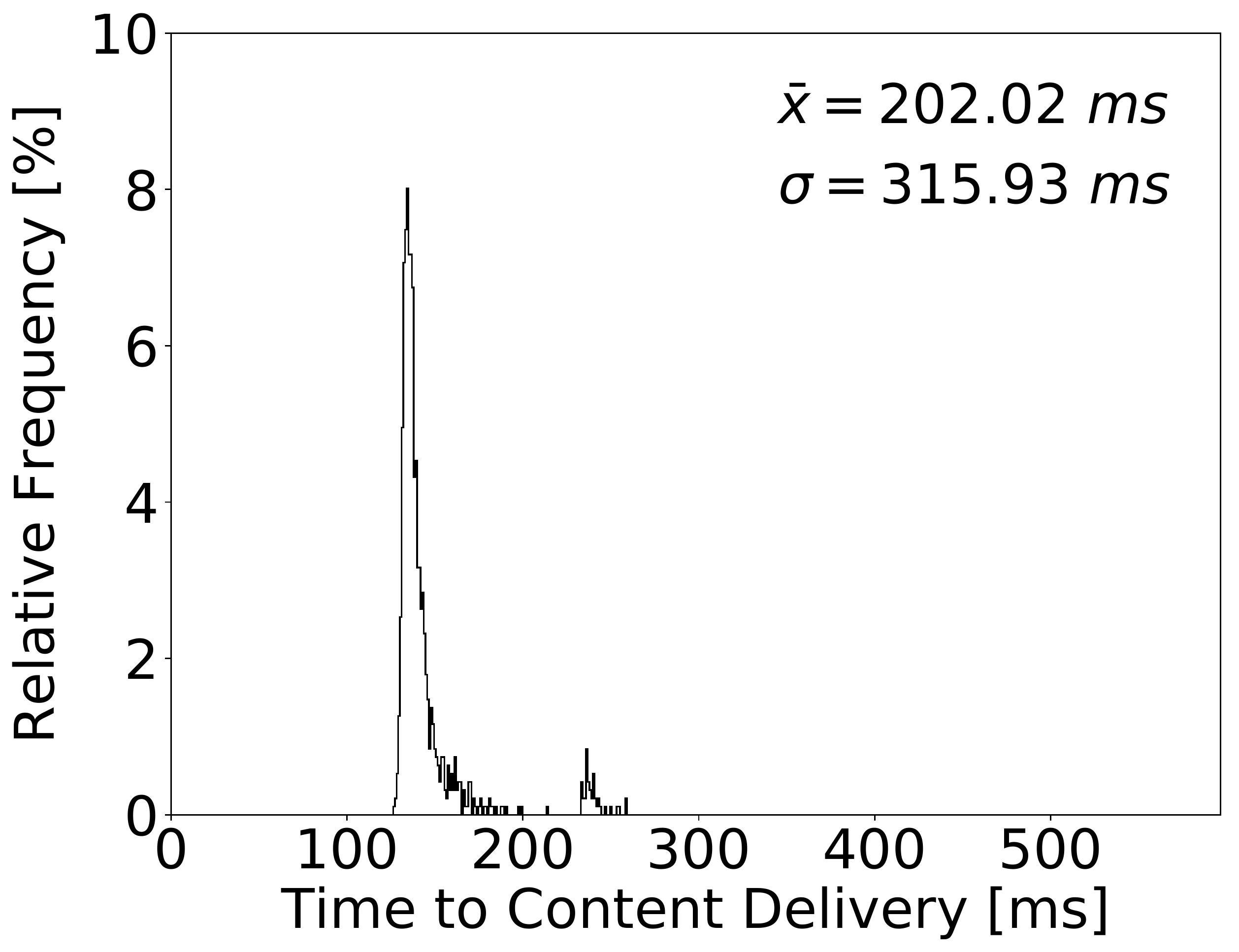}
    \label{fig:alerttime_paris}}
   \subfigure[Grenoble (ring) topology]{\includegraphics[width=0.66\columnwidth]{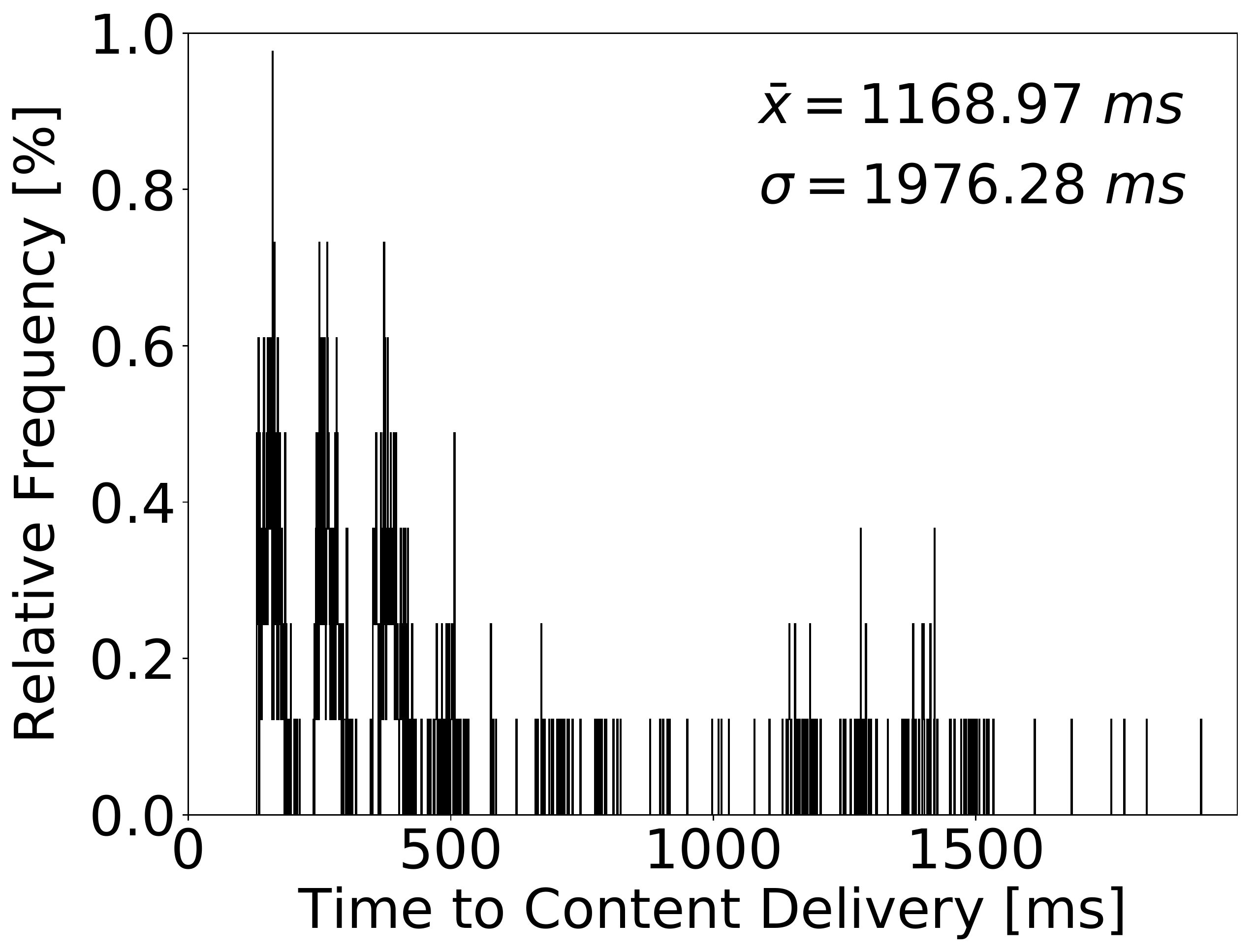}
    \label{fig:alerttime_grenoble-ring}}
   \subfigure[Grenoble topology]{\includegraphics[width=0.66\columnwidth]{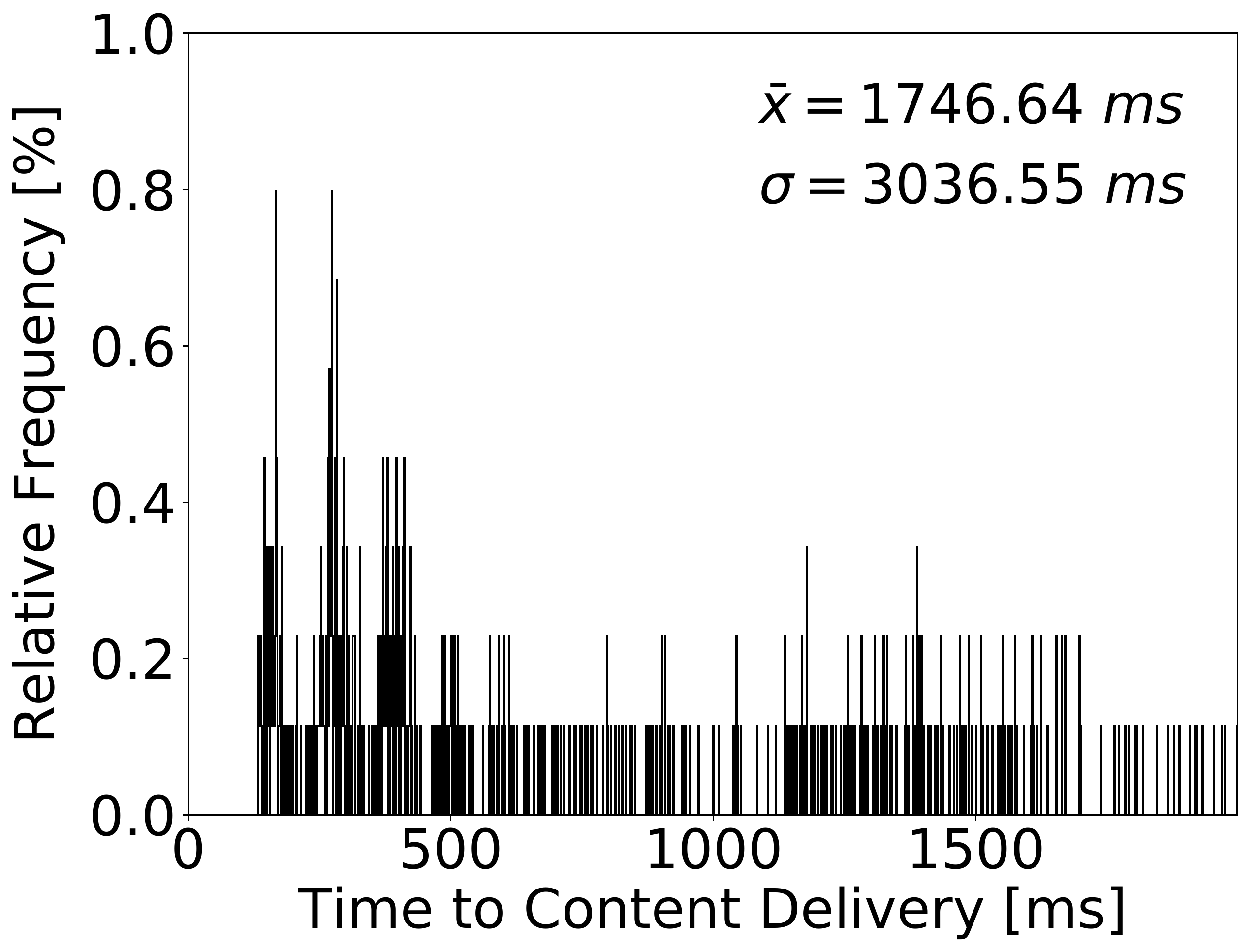}
    \label{fig:alerttime_grenoble}}
  \caption{Time to issue alerts}
\end{figure*} 

The experimental output for the three topologies are displayed in Figs. \ref{fig:alerttime_paris}, \ref{fig:alerttime_grenoble-ring}, and \ref{fig:alerttime_grenoble} respectively. As we might expect, blurring fluctuations have enhanced with only a few pronounced signatures of hops and the means increased slightly by the extended paths towards the subscribers. Notably, the single-hop testbed from Paris performed best under the extended communication load, whereas the full Grenoble testbed clearly runs at its limit. The latter can be easily explained by the many hop transitions required at Grenoble, each of which requires an additional packet exchange which potentially impacts on neighbors within radio range. 

Low power lossy networks that connect heavily constrained IoT nodes are known to be infeasible for such heavy load. We consider it therefore a success that a notable fraction of the content arrived at its receivers on within about 500 ms -- a timescale which is considered normal in multi-hop WPANs. To a certain degree, we account this for the robustness of our hopwise content publishing and replication protocol.


\section{Conclusions and Outlook}
\label{sec:c+o}

Future IoT networking is one of the most challenging use cases of the Internet today and a potential deployment regime of ICN. In this work, we revisited Information-Centric Networking in the IoT from a variety of perspectives and concluded that (a) publish-subscribe with named topics largely facilitates to manage the complexity of naming data, and (b) NDN without a push option for data has striking advantages for security and resilience in constrained  environments. We propose HoPP,  a lightweight publish-subscribe system that was implemented on RIOT and CCN-lite and experimentally evaluated on large, realistic testbeds. Our findings confirmed that constrained lossy networks can admit largely unforeseeable behaviour. Nevertheless, our approach turned out robust and resilient while performing well in the majority of experiments. 

In future work, we will enhance our implementation and work towards prototypic deployment in more intricate use cases. Prior to that, we will study mobility and disruption tolerance in closer detail using multi-proxy set-ups and content redundancy. Adding an analytic model that complements our understanding of the different protocol control loops will be  valuable for optimizing parameters and the overall performance.

\balance

\bibliographystyle{IEEEtran}
\bibliography{own,rfcs,ids,theory,complexity,layer2,internet,transport,overlay,hypermedia,vcoip,visualization,security,ngi,manet}

\begin{thebibliography}{10}
\providecommand{\url}[1]{#1}
\csname url@samestyle\endcsname
\providecommand{\newblock}{\relax}
\providecommand{\bibinfo}[2]{#2}
\providecommand{\BIBentrySTDinterwordspacing}{\spaceskip=0pt\relax}
\providecommand{\BIBentryALTinterwordstretchfactor}{4}
\providecommand{\BIBentryALTinterwordspacing}{\spaceskip=\fontdimen2\font plus
\BIBentryALTinterwordstretchfactor\fontdimen3\font minus
  \fontdimen4\font\relax}
\providecommand{\BIBforeignlanguage}[2]{{%
\expandafter\ifx\csname l@#1\endcsname\relax
\typeout{** WARNING: IEEEtran.bst: No hyphenation pattern has been}%
\typeout{** loaded for the language `#1'. Using the pattern for}%
\typeout{** the default language instead.}%
\else
\language=\csname l@#1\endcsname
\fi
#2}}
\providecommand{\BIBdecl}{\relax}
\BIBdecl

\bibitem{adiko-sind-12}
B.~Ahlgren, C.~Dannewitz, C.~Imbrenda, D.~Kutscher, and B.~Ohlman, ``{A Survey
  of Information-Centric Networking},'' \emph{IEEE Communications Magazine},
  vol.~50, no.~7, pp. 26--36, July 2012.

\bibitem{xvsft-sinr-14}
G.~Xylomenos, C.~N. Ververidis, V.~A. Siris, N.~Fotiou, C.~Tsilopoulos,
  X.~Vasilakos, K.~V. Katsaros, and G.~C. Polyzos, ``{A Survey of
  Information-Centric Networking Research},'' \emph{IEEE Communications Surveys
  and Tutorials}, vol.~16, no.~2, pp. 1024--1049, 2014.

\bibitem{lvt-psipp-10}
D.~Lagutin, K.~Visala, and S.~Tarkoma, ``{Publish/Subscribe for Internet: PSIRP
  Perspective},'' \emph{Future internet assembly}, vol.~84, 2010.

\bibitem{cpw-cpsni-11}
A.~Carzaniga, M.~Papalini, and A.~L. Wolf, ``{Content-based Publish/Subscribe
  Networking and Information-centric Networking},'' in \emph{Proc. of the ACM
  SIGCOMM WS on Information-centric Networking (ICN '11)}.\hskip 1em plus 0.5em
  minus 0.4em\relax New York, NY, USA: ACM, 2011, pp. 56--61.

\bibitem{cajfr-cecop-11}
J.~Chen, M.~Arumaithurai, L.~Jiao, X.~Fu, and K.~Ramakrishnan, ``{COPSS: An
  Efficient Content Oriented Publish/Subscribe System},'' in \emph{ACM/IEEE
  Symposium on Architectures for Networking and Communications Systems
  (ANCS'11)}.\hskip 1em plus 0.5em minus 0.4em\relax Los Alamitos, CA, USA:
  IEEE Computer Society, Oct. 2011, pp. 99--110.

\bibitem{jstp-nnc-09}
V.~Jacobson, D.~K. Smetters, J.~D. Thornton, and M.~F. Plass, ``{Networking
  Named Content},'' in \emph{5th Int. Conf. on emerging Networking Experiments
  and Technologies (ACM CoNEXT'09)}.\hskip 1em plus 0.5em minus 0.4em\relax New
  York, NY, USA: ACM, Dec. 2009, pp. 1--12.

\bibitem{bgnt-sieoc-13}
J.~Burke, P.~Gasti, N.~Nathan, and G.~Tsudik, ``{Securing Instrumented
  Environments over Content-Centric Networking: the Case of Lighting Control
  and NDN},'' in \emph{Computer Communications Workshops (INFOCOM WKSHPS), 2013
  IEEE Conference on}.\hskip 1em plus 0.5em minus 0.4em\relax IEEE, 2013, pp.
  394--398.

\bibitem{gkrao-dcm-10}
P.~Ginzboorg, T.~K{\"a}rkk{\"a}inen, A.~Ruotsalainen, M.~Andersson, and J.~Ott,
  ``{DTN Communication in a Mine},'' in \emph{2nd Extreme Workshop on
  Communications}.\hskip 1em plus 0.5em minus 0.4em\relax ACM, Sept. 2010.

\bibitem{gkslp-inii-17}
C.~G{\"u}ndogan, P.~Kietzmann, T.~C. Schmidt, M.~Lenders, H.~Petersen,
  M.~W{\"a}hlisch, M.~Frey, and F.~Shzu-Juraschek, ``{Information-Centric
  Networking for the Industrial IoT},'' in \emph{Proc. of 4th ACM Conference on
  Information-Centric Networking (ICN), Demo Session}.\hskip 1em plus 0.5em
  minus 0.4em\relax New York, NY, USA: ACM, September 2017, pp. 214--215.

\bibitem{olg-ccnte-10}
S.~Y. Oh, D.~Lau, and M.~Gerla, ``{Content Centric Networking in tactical and
  emergency MANETs},'' in \emph{2010 IFIP Wireless Days}.\hskip 1em plus 0.5em
  minus 0.4em\relax IEEE, Oct 2010, pp. 1--5.

\bibitem{bmhsw-icnie-14}
\BIBentryALTinterwordspacing
E.~Baccelli, C.~Mehlis, O.~Hahm, T.~C. Schmidt, and M.~W{\"a}hlisch,
  ``{Information Centric Networking in the IoT: Experiments with NDN in the
  Wild},'' in \emph{Proc. of 1st ACM Conf. on Information-Centric Networking
  (ICN-2014)}.\hskip 1em plus 0.5em minus 0.4em\relax New York: ACM, September
  2014, pp. 77--86. [Online]. Available:
  \url{http://dx.doi.org/10.1145/2660129.2660144}
\BIBentrySTDinterwordspacing

\bibitem{RFC-7476}
K.~Pentikousis, B.~Ohlman, D.~Corujo, G.~Boggia, G.~Tyson, E.~Davies,
  A.~Molinaro, and S.~Eum, ``{Information-Centric Networking: Baseline
  Scenarios},'' IETF, RFC 7476, March 2015.

\bibitem{wsv-bipmc-12}
\BIBentryALTinterwordspacing
M.~W{\"a}hlisch, T.~C. Schmidt, and M.~Vahlenkamp, ``{Bulk of Interest:
  Performance Measurement of Content-Centric Routing},'' in \emph{Proc. of ACM
  SIGCOMM, Poster Session}.\hskip 1em plus 0.5em minus 0.4em\relax New York:
  ACM, August 2012, pp. 99--100. [Online]. Available:
  \url{http://conferences.sigcomm.org/sigcomm/2012/paper/sigcomm/p99.pdf}
\BIBentrySTDinterwordspacing

\bibitem{gtuz-ddndn-13}
P.~Gasti, G.~Tsudik, E.~Uzun, and L.~Zhang, ``{DoS and DDoS in Named Data
  Networking},'' in \emph{Proc. of ICCCN}.\hskip 1em plus 0.5em minus
  0.4em\relax {IEEE}, 2013, pp. 1--7.

\bibitem{wsv-bdpts-13}
\BIBentryALTinterwordspacing
M.~W{\"a}hlisch, T.~C. Schmidt, and M.~Vahlenkamp, ``{Backscatter from the Data
  Plane -- Threats to Stability and Security in Information-Centric Network
  Infrastructure},'' \emph{Computer Networks}, vol.~57, no.~16, pp. 3192--3206,
  Nov. 2013. [Online]. Available:
  \url{http://dx.doi.org/10.1016/j.comnet.2013.07.009}
\BIBentrySTDinterwordspacing

\bibitem{sws-rcani-15}
S.~Al-Sheikh, M.~W{\"a}hlisch, and T.~C. Schmidt, ``{Revisiting Countermeasures
  Against NDN Interest Flooding},'' in \emph{2nd ACM Conference on
  Information-Centric Networking, Poster Session}, ser. ICN 2015.\hskip 1em
  plus 0.5em minus 0.4em\relax New York: ACM, Oct. 2015, pp. 195--196.

\bibitem{pf-britu-15}
G.~C. Polyzos and N.~Fotiou, ``{Building a reliable Internet of Things using
  Information-Centric Networking},'' \emph{Journal of Reliable Intelligent
  Environments}, vol.~1, no.~1, pp. 47--58, 2015.

\bibitem{sqvcg-simab-16}
J.~Suarez, J.~Quevedo, I.~Vidal, D.~Corujo, J.~Garcia-Reinoso, and R.~L.
  Aguiar, ``{A secure IoT management architecture based on Information-Centric
  Networking},'' \emph{Journal of Network and Computer Applications}, vol.~63,
  pp. 190 -- 204, 2016.

\bibitem{abcmr-inmcd-16}
M.~Amadeo, O.~Briante, C.~Campolo, A.~Molinaro, and G.~Ruggeri,
  ``{Information-centric networking for M2M communications: Design and
  deployment},'' \emph{Computer Communications}, vol. 89--90, pp. 105 -- 116,
  2016.

\bibitem{mwt-tucin-16}
B.~Mathieu, C.~Westphal, and P.~Truong, ``Towards the usage of ccn for iot
  networks,'' in \emph{Internet of Things (IoT) in 5G Mobile
  Technologies}.\hskip 1em plus 0.5em minus 0.4em\relax Springer, 2016, pp.
  3--24.

\bibitem{g-ainai-17}
J.~J. Garcia-Luna-Aceves, ``{ADN: An Information-Centric Networking
  Architecture for the Internet of Things},'' in \emph{Proc. of the 2nd
  International Conference on Internet-of-Things Design and Implementation},
  ser. IoTDI '17.\hskip 1em plus 0.5em minus 0.4em\relax ACM, 2017, pp. 27--36.

\bibitem{szsmb-avdir-17}
E.~M. Schooler, D.~Zage, J.~Sedayao, H.~Moustafa, A.~Brown, and M.~Ambrosin,
  ``{An Architectural Vision for a Data-Centric IoT: Rethinking Things, Trust
  and Clouds},'' in \emph{IEEE 37th Intern. Conference on Distributed Computing
  Systems (ICDCS)}.\hskip 1em plus 0.5em minus 0.4em\relax IEEE, June 2017, pp.
  1717--1728.

\bibitem{acim-icnis-15}
M.~Amadeo, C.~Campolo, A.~Iera, and A.~Molinaro, ``{Information Centric
  Networking in IoT scenarios: The case of a smart home},'' in \emph{2015 IEEE
  International Conference on Communications (ICC)}, June 2015, pp. 648--653.

\bibitem{srs-sndnt-15}
D.~Saxena, V.~Raychoudhury, and N.~SriMahathi, ``{SmartHealth-NDNoT: Named Data
  Network of Things for Healthcare Services},'' in
  \emph{MobileHealth@MobiHoc}.\hskip 1em plus 0.5em minus 0.4em\relax IEEE,
  2015, pp. 45--50.

\bibitem{habsw-itpla-16}
\BIBentryALTinterwordspacing
O.~Hahm, C.~Adjih, E.~Baccelli, T.~C. Schmidt, and M.~W{\"a}hlisch, ``{ICN over
  TSCH: Potentials for Link-Layer Adaptation in the IoT},'' in \emph{Proc. of
  3rd ACM Conf. on Information-Centric Networking (ICN 2016), Poster
  Session}.\hskip 1em plus 0.5em minus 0.4em\relax ACM, September 2016, pp.
  195---196, best Poster Award. [Online]. Available:
  \url{http://dx.doi.org/10.1145/2984356.2985226}
\BIBentrySTDinterwordspacing

\bibitem{ccn-lite}
\BIBentryALTinterwordspacing
``{CCN Lite: Lightweight implementation of the Content Centric Networking
  protocol},'' 2014. [Online]. Available: \url{http://ccn-lite.net}
\BIBentrySTDinterwordspacing

\bibitem{bhgws-rotoi-13}
E.~Baccelli, O.~Hahm, M.~G{\"u}nes, M.~W{\"a}hlisch, and T.~C. Schmidt, ``{RIOT
  OS: Towards an OS for the Internet of Things},'' in \emph{Proc. of the 32nd
  IEEE INFOCOM. Poster}.\hskip 1em plus 0.5em minus 0.4em\relax Piscataway, NJ,
  USA: IEEE Press, 2013.

\bibitem{dgv-clfos-04}
A.~Dunkels, B.~Gronvall, and T.~Voigt, ``Contiki---a lightweight and flexible
  operating system for tiny networked sensors,'' in \emph{Local Computer
  Networks, 2004. 29th Annual IEEE International Conference on}.\hskip 1em plus
  0.5em minus 0.4em\relax IEEE, 2004, pp. 455--462.

\bibitem{alw-defsc-16}
B.~Ahlgren, A.~Lindgren, and Y.~Wu, ``{Demo: Experimental Feasibility Study of
  CCN-lite on Contiki Motes for IoT Data Streams},'' in \emph{Proceedings of
  the 2016 conference on 3rd ACM Conference on Information-Centric
  Networking}.\hskip 1em plus 0.5em minus 0.4em\relax ACM, 2016, pp. 221--222.

\bibitem{saz-dinps-16}
W.~Shang, A.~Afanasyev, and L.~Zhang, ``{The Design and Implementation of the
  NDN Protocol Stack for RIOT-OS},'' in \emph{Proc. of IEEE GLOBECOM
  2016}.\hskip 1em plus 0.5em minus 0.4em\relax Washington, DC, USA: IEEE,
  2016, pp. 1--6.

\bibitem{RFC-7927}
D.~Kutscher, S.~Eum, K.~Pentikousis, I.~Psaras, D.~Corujo, D.~Saucez,
  T.~Schmidt, and M.~Waehlisch, ``{Information-Centric Networking (ICN)
  Research Challenges},'' IETF, RFC 7927, July 2016.

\bibitem{acim-ndnia-14}
M.~Amadeo, C.~Campolo, A.~Iera, and A.~Molinaro, ``{Named data networking for
  IoT: An architectural perspective},'' in \emph{2014 European Conference on
  Networks and Communications (EuCNC)}.\hskip 1em plus 0.5em minus 0.4em\relax
  IEEE, June 2014, pp. 1--5.

\bibitem{draft-ravi-icnrg-ccn-notification}
R.~Ravindran, A.~Chakraborti, S.~Amin, and J.~Chen, ``{Support for
  Notifications in CCN},'' IETF, Internet-Draft -- work in progress~01, July
  2017.

\bibitem{kgshw-nnmam-17}
P.~Kietzmann, C.~G{\"u}ndogan, T.~C. Schmidt, O.~Hahm, and M.~W{\"a}hlisch,
  ``{The Need for a Name to MAC Address Mapping in NDN: Towards Quantifying the
  Resource Gain},'' in \emph{Proc. of 4th ACM Conference on Information-Centric
  Networking (ICN)}.\hskip 1em plus 0.5em minus 0.4em\relax New York, NY, USA:
  ACM, September 2017, pp. 36--42.

\bibitem{wsv-lpwds-13}
M.~W{\"a}hlisch, T.~C. Schmidt, and M.~Vahlenkamp, ``{Lessons from the Past:
  Why Data-driven States Harm Future Information-Centric Networking},'' in
  \emph{Proc. of IFIP Networking}.\hskip 1em plus 0.5em minus 0.4em\relax
  Piscataway, NJ, USA: IEEE Press, 2013.

\bibitem{g-ncric-14}
J.~Garcia-Luna-Aceves, ``{Name-based Content Routing in Information Centric
  Networks Using Distance Information},'' in \emph{1st ACM Conference on
  Information-Centric Networking}, ser. ACM-ICN '14.\hskip 1em plus 0.5em minus
  0.4em\relax ACM, 2014, pp. 7--16.

\bibitem{hg-nanlr-15}
E.~Hemmati and J.~Garcia-Luna-Aceves, ``{A New Approach to Name-Based
  Link-State Routing for Information-Centric Networks},'' in \emph{2Nd ACM
  Conference on Information-Centric Networking}, ser. ACM-ICN '15.\hskip 1em
  plus 0.5em minus 0.4em\relax ACM, 2015, pp. 29--38.

\bibitem{wboks-puscd-15}
L.~Wang, S.~Bayhan, J.~Ott, J.~Kangasharju, A.~Sathiaseelan, and J.~Crowcroft,
  ``{Pro-Diluvian: Understanding Scoped-Flooding for Content Discovery in
  Information-Centric Networking},'' in \emph{2Nd ACM Conference on
  Information-Centric Networking}, ser. ACM-ICN '15.\hskip 1em plus 0.5em minus
  0.4em\relax New York, NY, USA: ACM, 2015, pp. 9--18.

\bibitem{za-lcdds-13}
Z.~Zhu and A.~Afanasyev, ``{Let's ChronoSync: Decentralized Dataset State
  Synchronization in Named Data Networking},'' in \emph{Proc. of the 21st IEEE
  International Conference on Network Protocols (ICNP 2013)}, 2013.

\bibitem{fbc-snibf-15}
W.~Fu, H.~Ben~Abraham, and P.~Crowley, ``{Synchronizing Namespaces with
  Invertible Bloom Filters},'' in \emph{11th ACM/IEEE Symposium on
  Architectures for Networking and Communications Systems}, ser. ANCS
  '15.\hskip 1em plus 0.5em minus 0.4em\relax IEEE Computer Society, 2015, pp.
  123--134.

\bibitem{zlw-pespn-16}
\BIBentryALTinterwordspacing
M.~Zhang, V.~Lehman, and L.~Wang, ``{PartialSync: Efficient Synchronization of
  a Partial Namespace in NDN},'' no. NDN-0039-1, Jun. 2016. [Online].
  Available:
  \url{https://named-data.net/wp-content/uploads/2016/06/ndn-0039-1-partial-sync.pdf}
\BIBentrySTDinterwordspacing

\bibitem{haazz-nnlsr-13}
M.~Hoque, S.~O. Amin, A.~Alyyan, B.~Zhang, L.~Zhang, and L.~Wang, ``{NLSR:
  Named-data Link State Routing Protocol},'' in \emph{3rd ACM SIGCOMM Workshop
  on Information-centric Networking}, ser. ICN '13.\hskip 1em plus 0.5em minus
  0.4em\relax New York, NY, USA: ACM, 2013, pp. 15--20.

\bibitem{swbw-lcnhp-16}
T.~C. Schmidt, S.~W{\"o}lke, N.~Berg, and M.~W{\"a}hlisch, ``{Let's Collect
  Names: How PANINI Limits FIB Tables in Name Based Routing},'' in \emph{Proc.
  of 15th IFIP Networking Conference}.\hskip 1em plus 0.5em minus 0.4em\relax
  Piscataway, NJ, USA: IEEE Press, May 2016, pp. 458--466.

\bibitem{draft-ietf-core-senml}
C.~Jennings, Z.~Shelby, J.~Arkko, A.~Keranen, and C.~Bormann, ``{Media Types
  for Sensor Measurement Lists (SenML)},'' IETF, Internet-Draft -- work in
  progress~10, July 2017.

\bibitem{l-rhm-89}
G.~P. Landow, ``{The rhetoric of hypermedia: Some rules for authors},''
  \emph{Journ. of Comp. in Higher Education}, vol.~1, no.~1, pp. 39--64, 1989.

\bibitem{yamwz-csfp-12}
C.~Yi, A.~Afanasyev, I.~Moiseenko, L.~Wang, B.~Zhang, and L.~Zhang, ``{A Case
  for Stateful Forwarding Plane},'' PARC, Tech. Rep. NDN-0002, July 2012.

\bibitem{yaawz-rrndn-14}
C.~Yi, J.~Abraham, A.~Afanasyev, L.~Wang, B.~Zhang, and L.~Zhang, ``{On the
  Role of Routing in Named Data Networking},'' in \emph{Proceedings of the 1st
  ACM Conference on Information-Centric Networking}, ser. ACM-ICN '14.\hskip
  1em plus 0.5em minus 0.4em\relax New York, NY, USA: ACM, 2014, pp. 27--36.

\bibitem{gm-lfpcn-16}
J.~J. Garcia-Luna-Aceves and M.~Mirzazad-Barijough, ``A light-weight forwarding
  plane for content-centric networks,'' in \emph{2016 International Conference
  on Computing, Networking and Communications (ICNC)}.\hskip 1em plus 0.5em
  minus 0.4em\relax IEEE, Feb 2016, pp. 1--7.

\bibitem{RFC-4601}
B.~Fenner, M.~Handley, H.~Holbrook, and I.~Kouvelas, ``{Protocol Independent
  Multicast - Sparse Mode (PIM-SM): Protocol Specification (Revised)},'' IETF,
  RFC 4601, August 2006.

\bibitem{swbw-panii-15}
T.~C. Schmidt, S.~W{\"o}lke, N.~Berg, and M.~W{\"a}hlisch, ``{Partial Adaptive
  Name Information in ICN: PANINI Routing Limits FIB Table Sizes},'' in
  \emph{2nd ACM Conference on Information-Centric Networking, Poster Session},
  ser. ICN 2015.\hskip 1em plus 0.5em minus 0.4em\relax New York: ACM, Oct.
  2015, pp. 193--194.

\bibitem{tscrm-smin-13}
G.~Tyson, N.~Sastry, R.~Cuevas, I.~Rimac, and A.~Mauthe, ``{A Survey of
  Mobility in Information-centric Networks},'' \emph{Commun. ACM}, vol.~56,
  no.~12, pp. 90--98, Dec. 2013.

\bibitem{wakvw-rtidu-12}
L.~Wang, A.~Afanasyev, R.~Kuntz, R.~Vuyyuru, R.~Wakikawa, and L.~Zhang,
  ``{Rapid Traffic Information Dissemination Using Named Data},'' in
  \emph{Proc. of 1st ACM Workshop on Emerging Name-Oriented Mobile Networking
  Design - Architecture, Algorithms, and Applications (NoM)}.\hskip 1em plus
  0.5em minus 0.4em\relax New York, NY, USA: ACM, 2012, pp. 7--12.

\bibitem{gpwpv-hpvin-13}
G.~Grassi, D.~Pesavento, L.~Wang, G.~Pau, R.~Vuyyuru, R.~Wakikawa, and
  L.~Zhang, ``{ACM HotMobile 2013 Poster: Vehicular Inter-networking via Named
  Data},'' \emph{SIGMOBILE Mob. Comput. Commun. Rev.}, vol.~17, no.~3, pp.
  23--24, November 2013.

\bibitem{RFC-6550}
T.~Winter, P.~Thubert, A.~Brandt, J.~Hui, R.~Kelsey, P.~Levis, K.~Pister,
  R.~Struik, J.~Vasseur, and R.~Alexander, ``{RPL: IPv6 Routing Protocol for
  Low-Power and Lossy Networks},'' IETF, RFC 6550, March 2012.

\bibitem{ivlso-14}
``{IoT-LAB: a very large scale open testbed},'' \url{https//www.iot-lab.info/},
  2015.

\end{thebibliography}

\end{document}